\documentclass[
draftclsnofoot,
onecolumn,
]{IEEEtran}

\usepackage{float}
\usepackage[utf8]{inputenc} \usepackage[T1]{fontenc}    \usepackage{hyperref}       

\usepackage{url}            \usepackage{booktabs}       \usepackage{amsmath,amsfonts}       \usepackage{nicefrac}       \usepackage{microtype}      \usepackage{array}
\RequirePackage{etexcmds}
\usepackage{flushend}
\usepackage{color}
\usepackage{graphicx}
\usepackage{epstopdf}
\usepackage{subcaption}
\usepackage{multirow}
\usepackage[table,xcdraw]{xcolor}
\usepackage{amsmath,stackrel}
\usepackage{amsfonts}
\usepackage{mathalfa}
\usepackage{amssymb,mathtools}
\usepackage{amscd}
\usepackage{multirow}
\usepackage{makecell}
\usepackage{booktabs}
\usepackage[ruled,vlined,commentsnumbered,linesnumbered,lined,boxed]{algorithm2e}
\usepackage{algorithmicx}
\usepackage{algpseudocode}

\usepackage{ifthen} \usepackage{xifthen} 

\usepackage{xpatch}
\usepackage{blindtext}

\usepackage[colorinlistoftodos,prependcaption,textsize=tiny]{todonotes}

\usepackage{listings}
\usepackage{paralist}
\usepackage{enumerate}
\usepackage[shortlabels]{enumitem}
\newlist{abbrv}{itemize}{1}
\setlist[abbrv,1]{label=,labelwidth=1in,align=parleft,itemsep=0.1\baselineskip,leftmargin=!, font=\normalfont\large}

\usepackage{pgf}
\usepackage{tikz}
\usetikzlibrary{decorations.text}
\usetikzlibrary{backgrounds,shapes} \usetikzlibrary{arrows,automata}

\def\Cwnd{\textbf{\textit{Cwnd }}}

\SetKwProg{Fn}{Function}{}{}

\newcommand*{\affmark}[1][*]{\textsuperscript{#1}}

\newcommand\scheme{{\textit{{RWNDQ }}}} 

\title{Design and Implementation of Fair Congestion Control for Data Centers Networks\thanks{A preliminary versions of this work were published in part in IEEE CloudNet \cite{Ahmed-CLOUDNET-2015} and IEEE IPCCC \cite{Ahmed-IPCCC-2015}.}}

\author{Ahmed M. Abdelmoniem\affmark[1,2] \quad and \quad Brahim Bensaou\affmark[2]\\
\affmark[1]Assuit University, Egypt \quad \affmark[2]HKUST, Hong Kong}

\begin{document}

\bstctlcite{IEEEexample:BSTcontrol}

\maketitle

\begin{abstract}
In data centers, the nature of the composite bursty traffic along with the small bandwidth-delay product and switch buffers lead to several congestion problems that are not handled well by traditional congestion control mechanisms such as TCP. Existing work try to address the problem by modifying TCP to suit the operational nature of data centers. This is practically feasible in private settings, however, in public environments, such modifications are prohibited. Therefore, in this work, we design a simple switch-based queue management to deal with such congestion issues adequately. This approach entails no modification to the TCP sender and receiver algorithms which enables easy and seamless deployment in public data centers. We present theoretical analysis to show the stability and effectiveness of the scheme. We also present, three different real implementations (as a Linux kernel module and as an added feature to OpenvSwitch) and give numerical results from both NS-2 simulation and experiments of real deployment in a small test-bed cluster to show its effectiveness in achieving high throughput overall, a good fairness and short flow completion times for delay-sensitive flows.
\end{abstract}

\section{Introduction}

In this work, we focus on the ongoing tug of war between mice and elephants in data centers and to the many works proposed in the literature to reconcile their conflicting requirement in particular \cite{GuoLiang2001}. Numerous congestion control mechanisms have been proposed for the Internet, high speed WANs, lossy wireless networks and data centers, and the interested reader may refer to the following surveys and references therein for a comprehensive coverage~\cite{TCPsurvey, TCPfriendlysurvey, TCPwireless, TCPdatacenter}. Each algorithm aims to improve the way TCP reacts to congestion; some propose conservative and friendly adjustments, while others propose aggressive ones to quickly fill the communication pipe and improve link utilization. In contrast, we follow another intuitive approach that relies on the well-established \textbf{``Flow Control"} algorithm in TCP to achieve congestion control. The flow control algorithm is a historically older part of TCP than the congestion control algorithm \cite{RFCTCP}. Hence, we adopt a different view which focuses on one of the basic building blocks of the TCP protocol (i.e., Flow Control) and leverage the fact that it can control the sending rate of the sources. 

As discussed earlier, the co-existence of various flows with different performance requirements ranging from synchronous mice to bulky elephant flows, poses serious challenges to TCP. In addition, the high bandwidth, low latency and small buffered switched network used in data centers is the wrong recipe for TCP that was geared for the Internet with its relatively low bandwidth, high latency, and large buffered routers. This mismatch between TCP's design and data center's characteristics~\cite{Chen2009, Alizadeh2010, Ahmed-ANNALS-2017, Ahmed-ICC-2017, Ahmed-LCN-2017} leads to complex non-conventional congestion events which are typically not observed on the Internet. These events cannot simply be inferred from packet losses or duplicate ACKs, and hence they require special treatment to address. As a result, the most common TCP issues found in data centers are incast congestion and buffer-bloating which is defined as follows:
\begin{enumerate}
\item Incast congestion, where many correlated mice flows converge onto the same congested output port of a switch over a short period of time; and 
\item Queue-buildup/Buffer-Bloating that occurs as the normal grab-all-bandwidth behavior of TCP which results in elephant flows occupying most of the buffer space. In this case, mice flows fail to grab their share of buffer space and may see significantly long queuing delays.
\end{enumerate}
These two problems may occurs simultaneously due to the co-existence of flows of different nature and requirements in the same network.  For instance, buffer-bloating which is mainly due to long-lived elephants consumes scarce buffers of the bottleneck shared with partition/aggregate mice flows. When many such mice flows arrive almost synchronized (which is common for such type of applications), they may experience packet losses that are often recovered via retransmission after a long timeout. Since the common resource here is the buffer, we need to consider techniques that regulate the sending rate and share the buffer taken by each active flows in a fair proportion. 

Therefore, in this work, we take a flow-aware approach similar to traditional flow-based systems such as ATM available bit rate service (ATM-ABR) \cite{ATM-ABR} or its Internet counterpart XCP \cite{Katabi2002}. The challenge that arises however is how to deploy such flow-awareness in the flow-agnostic and aversive IP environment without modifying the TCP sender nor receiver. This disqualifies XCP, as it is a clean-slate redesign that requires not only changes to the routers but also to the sender and receiver. To achieve our goal, the switch must be able to track the number of active TCP flows per queue (instead of full TCP state), calculate a fair share for each flow that traverses the output queue, and must have the means to convey this fair share back to the traffic sources. We accomplish this via simple means as follows:
\begin{inparaenum}[i)]
	\item using a simple counter (i.e., register), we track the number of active flows for each output queue in the switch; and
	\item by modifying the switch software, we enable rewriting the flow's fair share in the receiver window field of the TCP header as a means to enforcing a sending rate at the TCP source.
\end{inparaenum}
This can be interpreted as if the switch device is establishing flow control to regulate sender rates to operate at a certain buffer occupancy. TCP flow control, being a fundamental part of any TCP incarnation, including XCP and DCTCP, our proposed mechanism would fit well without any change to TCP endpoints (i.e., guest VMs).

To achieve this goal, we propose a simple switch-based equal share allocation algorithm called \scheme ~\cite{Ahmed-CLOUDNET-2015,Ahmed-IPCCC-2015} that does not require any modification to the TCP stack in the guest VM. We mathematically model the \scheme dynamics and show its convergence. Then, we evaluate the performance of the switch scheme via simulation and real testbed experiments.\footnote{To help interested readers reproduce our results and for openness, we make the code and scripts of our implementations, simulations, experiments available online at the following link: \underline{\textbf{\url{http://github.com/ahmedcs/RWNDQ}}}.}

The remainder of this paper is organized as follows, we first give a brief overview of the background and problems in Section~\ref{sec:problem}. Then, we discuss our proposed methodology and present the proposed switch queue management algorithm ``\scheme" in Section~\ref{sec:method}. We further develop a simple analytical model of \scheme to study its convergence and stability in Section~\ref{sec:convstab}. Then we evaluate its performance via simulation and compare it to the alternative approaches in Section~\ref{sec:sim}. We present the testbed experiments of the Linux kernel, Open vSwitch and the hardware prototypes of \scheme in Sections~\ref{sec:exp}. Finally, we give the related work in Section~\ref{sec:related} and summarize in Section~\ref{sec:conclude}.

\section{Background}\label{sec:problem}

In this section, we give the background information about the unfairness among TCP flows of the same or different variant. Then, we show that the tug of war between mice and elephants has been an issue for all sorts of networks including the Internet. To solve this problem without modifying the TCP, we leverage the built-in TCP flow control and a queue management scheme, hence we give a brief overview on TCP flow control and the role of queue management in the network.

\subsection{The Intra-Protocol Unfairness in Data Centers}

The coexistence of TCP flows from the same variant should normally result in an ideal fair competition where each flow grabs an equal share of the bottleneck link capacity. This fairness is also known as the Round-Trip Time (RTT) fairness because the condition to reach such fairness is dependent on the fairness of the RTT (because the throughput is inversely proportional to the RTT). Moreover, the design of TCP for the Internet targets the goal of achieving long-term fairness among competing flows, and this goal has been inherited by TCP deployments in current data center networks. The problem with this goal is that short-lived TCP flows do not last enough to reach the steady state where they can obtain their fairness. Therefore, they would benefit more if short-term fairness is achieved instead. With the current TCP design, it is hard to achieve this goal for both the long delay moderate speed networks (e.g., The Internet) and the low delay high speed networks (e.g., data centers). This means that long-lived flows that last longer (i.e., elephants) are in a more advantageous position over short-lived flows that last for just few RTTs (i.e., mice). This problem has been reported and studied more than a decade ago in the Internet~\cite{GuoLiang2001} and solutions based on flow classification and prioritization have been proposed~\cite{Matta2000, Yilmaz2001}. However, due to the complexity and non-practicality of these solutions, they could not see wide adoption from the networking community.

\subsection{The TCP Flow Control mechanism}

\begin{figure}[!ht]
	\centering
        \includegraphics[height=6cm, width=0.95\textwidth]{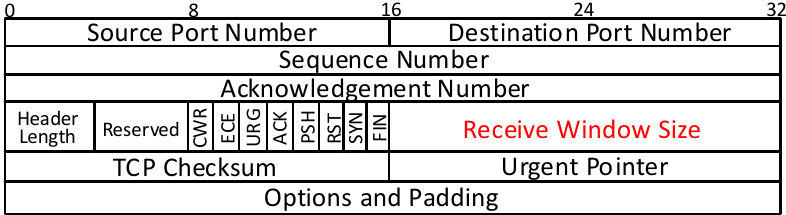}
         \caption{Transmission Control Protocol (TCP) headers. Receive window field is shown in red.}
        \label{fig:RWNDQTCPheader}				
\end{figure}

The main goal for flow control is for TCP to ensure that the sender will not overflow the buffer of the receiver. Normally, during TCP connection establishment, each TCP end-point reserves a buffer space for receiving incoming segments from its peer. The outgoing segments waiting for transmission or arriving segments waiting to be consumed by the application are stored in the send buffer or receive buffer, respectively. The reason is that TCP is a reliable protocol and it has to ensure the \textbf{``orderly delivery"} of data between the two communicating entities. In some networks, packets may arrive out of order or may be lost and/or some applications may be slow in consuming the incoming segments. Hence, these segments need to be buffered until the out-of-order data arrive or the slow application consumes the buffered segments. To prevent receive buffers from overflowing, TCP provides means for the receiver to pace the sender by controlling the extra amount of data that can sent by the sender. This is achieved by returning a permissible ``window" of bytes with every ACK. As shown in \figurename~\ref{fig:RWNDQTCPheader}, TCP segment headers comprises a field named ``receiver window size'' that serves the purpose of signaling the allowed number of bytes that the sender may transmit without overflowing the receive buffer~\cite{RFCTCP}.

\subsection{Relationship between Congestion Control and Flow Control}

TCP employs both a flow control and a congestion control mechanism. Even though they are considered two different mechanisms for different purposes, they are inter-related and both can be a limiting factor on the source sending rate. Generally speaking, the purpose of each is as follows:
\begin{itemize}
\item \textbf{Flow Control:} adjusts the sending rate of the traffic source to the available buffer space and processing speed of its peer.
\item \textbf{Congestion Control:} adapts the source rate of source to the current congestion state perceived from the network.
\end{itemize}
At any instant of time, any TCP flow in the network is limited by the either the remaining buffer space of its peer or the current value of its unused congestion window. TCP sets the relation between the congestion and receive window as follows: Swnd$  = min(\Cwnd , $Rwnd$)$. The TCP sending window is always the minimum among the current value of congestion window \Cwnd vs the current value of the receive window $Rwnd$ retrieved from the last ACK coming from its peer.

\subsection{The Role of Active Queue Management}

\begin{figure}[!ht]
	\centering
        \includegraphics[height=7cm, width=0.8\textwidth]{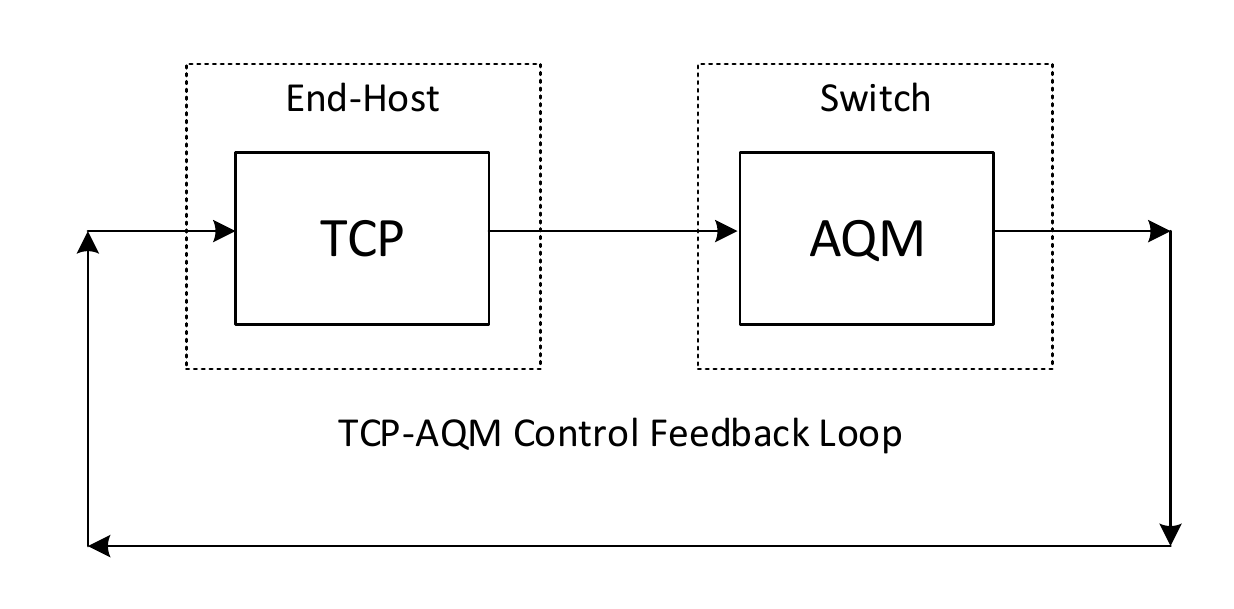}
         \caption{TCP and Active Queue Management Control Feedback Loop.}
        \label{fig:TCPAQM}				
\end{figure}

Active Queue Management (AQM) algorithms are deployed in switching devices to help in controlling congestion by continuously monitor the state of the output queues and taking an active role in relieving congestion if needed. Typically the instantaneous (or average) queue size, arrival rate and/or departure rate are estimated and whenever their value exceeds a certain threshold the algorithm infers congestion on the link. Upon congestion, either, the algorithm proactively drops packets as a form of implicit congestion notification to the sources or it sends explicit congestion notifications signals to the sources to adjust their sending rates. As typical example is the so called RED \cite{Floyd1993} with Explicit Congestion Notification (ECN). \figurename~\ref{fig:TCPAQM} shows the relation between TCP and AQM where both algorithms form a closed loop feedback control. In this control loop, the TCP sending rate affects the size of the queue in the switch and the AQM in turn acts as a control element which sends feedback signals back to TCP to regulate its sending rate.

\section{The Proposed Methodology}
\label{sec:method}

TCP is a full-duplex protocol where the two receiving end-points allocate a receiving buffer space to enable the flow control operations. To this end, the receiver of one direction sends back to the sender ACK packets that include in the header, in the 16 bit "Receive Window" field ($Rwnd$), the currently available buffer space. In high speed networks, to prevent the protocol from becoming a bottleneck, the receiver window scaling option of TCP is invoked. In this option, the two end-points agree a priori on a number ($n$) of bits that indicates the multiplicative factor ($2^n$) applied to the value of $Rwnd$ to calculate the actual receiver window value. This allows for receiver window values that can exceed the maximum of 64 KBytes, allowed by the 16 bits long $Rwnd$ field, to reach values of at up to 1 GByte for a scaling value of $n=14$. RFC1323~\cite{RFC1323} states that the window scaling option can either be set on all packets or only during the connection establishment phase. Linux TCP implementation takes the latter approach.

In our approach we propose to overwrite $Rwnd$ and the scaling option value $n$ in the ACKs to indicate the bottleneck fair share of the bandwidth and buffer available for each flow sharing the same output port. As the ACKs from the receiver traverse the switches in the reverse path towards the sender, each switch examines the ACK and modifies the $Rwnd$ value after adjusting it using the window scaling value if necessary. More specifically, at each switch along the end-to-end path:

\small
\begin{enumerate}
\item The available buffer space (i.e., queue occupancy) of each switch port is continuously sampled. 
\item Based on the number of ongoing flows and the drift from the target queue occupancy, a local window value, reflecting the per-flow fair share, is calculated. 
\item The $Rwnd$ field of the incoming ACK is rewritten with the local window only if the latter is smaller than the current value of the receive window field.
\item As the ACKs travel across the switches along the reverse path, $Rwnd$ is updated to reflect the fair share of bytes on the bottleneck for the flow.
\item Since the TCP sending window (or $Swnd$) is limited by the minimum of the current congestion window (or $Cwnd$) and the receiver window $Rwnd$, it is expected in steady state that the sender window will be limited by the receiver window which now reflects the bottleneck link fair share of bytes rather than the remaining bytes in the receive buffer. 
\end{enumerate}
\normalsize

\subsection{System Design and Algorithms}

\begin{table}[htbp]
	\caption{Variables and Parameters used in \scheme Algorithm \ref{alg:rwndq}}
	\centering
\begin{tabular}{|c|c|}
		\hline
		\textbf{Parameter name} 	& \textbf{Description} \\\hline
		$T$ 			& Timeout value for each window increment interval \\\hline
		$M$ 			& Number of increment intervals to wait for an update\\\hline
		$B$ 	         & Buffer size on the data path \\\hline
		$\alpha$ 		& Target level of queue occupancy	\\\hline\hline
		Variable name 	& Description \\\hline
		$Rwnd$ 		& Local receive window value for all flows\\\hline
		$\beta$     	& Number of current ongoing flows\\\hline
		$\gamma$ 	& Window increments of one update interval\\\hline
		$\Gamma$ 	& Counter of the number increments\\\hline
		$Q$ 		& Current output queue length in bytes  \\\hline
		$\kappa$    & The drift of $Q$ from the target $\alpha B$ \\\hline
		$P$		         & A packet\\\hline
		$Rwnd(P)$	& The value of receive window in TCP header\\\hline
		$Reserved(P)$	& The value of reserved bits in TCP header\\\hline
		$slow\_start$ & The current state of slow-start flag\\\hline
		\end{tabular}
\label{tab:varAlgo1}
\end{table}

The main variables and parameters used in \scheme algorithm are described in Table~\ref{tab:varAlgo1}. Note that $T$, $M$ and $\alpha$ are parameters of the algorithm that can be chosen by the DCN administrator. \scheme shown in Algorithm \ref{alg:rwndq} is event-driven and runs on the switch to respond to two major events: packet arrivals, and timer-based local window update events.  

\textbf{Upon a packet arrival:} the algorithm updates the maximum packet size seen so far. If this is the first flow to arrive at this port, then the current window is initially set to the target queue worth of bytes then \scheme (optionally) enters the slow-start phase to start probing for the effective window size. This is because initially the end-to-end bandwidth-delay product is unknown to the switch and hence the available bandwidth has to be probed. Subsequently, for each new flow, the current window is divided equally among all flows. If the ACK bit is set, the receive window field $Rwnd(P)$ is rescaled by the scale factor. Then the rescaled receive window $Rwnd(P)<<scale$ is compared to the current local window value $Rwnd$ of the ingress queue . If $Rwnd$ is smaller than $Rwnd(P)$, then this packet is updated with the current local window $Rwnd$ after being scaled by the scale factor (i.e., $Rwnd>>scale$). Note that, the scale factor can be encoded into the reserved bits field (i.e., $Reserved(P)$) by a hypervisor-level shim-layer running on the host physical machine.

\textbf{Upon window update timer expiry:} $\kappa$ is calculated to track the deviation of the current queue length from the target. The ratio controls the fraction of the Maximum Segment Size (MSS) added or subtracted from the current value of $\gamma$. After a number $\Gamma$ of such updates, the current value of the local per-queue window is updated. Then, if slow start is active, \scheme adds two MSS to the window, otherwise it adds the current value of $\gamma$ to the per-queue local window $Rwnd$. Notice that the value of the window increment is updated $M$ times before it is added to the actual value of TCP receive window $Rwnd(P)$ that is conveyed to the TCP sender. This enables a highly accurate estimate of the increment, while keeping the number of receive window field $Rwnd(P)$ rewrites in the packet header reasonable.

The design of \scheme enables it to maintain a very low loss rate and to leave enough buffer space to absorb sudden traffic bursts while keeping links highly utilized. Therefore, it is appealing for handling the co-existence of mice and elephants. \scheme adopts a proportional increase, proportional decrease approach that is the window is shrunk or expanded in proportion to the amount by which the queue occupancy is above or below the target queue level. Furthermore, the increase or decrease amount is equally divided among all ongoing flows. Initially and whenever the number of ongoing flows drops to zero, the algorithm goes into the slow start mode, where the window is incremented by two MSS after the end of each update period. When the queue exceeds the target, the algorithm goes into congestion avoidance and the window is decremented increment proportional to the difference between the current queue occupancy and the target queue occupancy. 

\begin{algorithm}[h]
 \caption{Switch-based Equal Fair-Share AQM (\scheme) Algorithm}
\label{alg:rwndq}
\Fn{Packet Departure Event Handler ($P$)}
{	
	\lIf{$MSS \leq TCPSize(P)$}
	{
               $MSS \leftarrow TCP\_Size(P)$
     }
	\If{$SYNACK(P)$}
	{
	         \lIf{$\beta \leq 0$}
	            {
					$Rwnd \leftarrow \alpha \times B$
		     	}			
		     \Else
		     {
			  $Rwnd \leftarrow Rwnd \times \frac{\beta}{\beta+1}$\;
		     }
		    $\beta \leftarrow \beta + 1$\;
     }
	\If{$FIN(P)$}
	{	
		 $\beta \leftarrow \beta - 1$\;
		\lIf{$\beta \geq 0$}
		{
			$Rwnd \leftarrow Rwnd \times \frac{\beta+1}{\beta}$
		}				   
		\Else
		{
			  $Rwnd \leftarrow \alpha \times B$\; 
			  $slow\_start \leftarrow True$\;			  
		}
	}
	\If{$ACK(P)$ \&\& ($Rwnd \leq Rwnd(P)<<Reserved(P)$)}
	{
		$Rwnd(P) \leftarrow Rwnd>>Reserved(P)$\;
	}
}
\Fn{Window Update Timer}
{
          $\kappa \leftarrow 1 - \frac{Q}{ B \times \alpha}$\;
          $\gamma\ \leftarrow \gamma  + \frac{\kappa \times MSS}{M}$\; 
          $\Gamma \leftarrow \Gamma + 1$\;
	      \If{$\Gamma == M$}
	      {
			 \lIf{$slow\_start == True$}
			 {
				$Rwnd \leftarrow Rwnd  + 2 \times MSS$
			}				
	        \lElse
	        {
				$Rwnd \leftarrow Rwnd  + \frac{\gamma}{\beta}$
	        }
	        \lIf{$Q \geq \alpha * B$}
			{
	                    $slowstart \leftarrow False$
			}
		    $\gamma \leftarrow 0$; $\Gamma \leftarrow 0$\;
	    }
}
\end{algorithm}
\normalsize

\subsection{Practical Aspects of The System}
\label{subsec:pracrwndq}

\textbf{Flow Tracking: } In principle \scheme is very effective in solving the problem of congestion, and actually avoiding it outright when there is no sudden traffic surge. However, to enable its practical deployment, 
\begin{inparaenum}[\itshape i)\upshape]
\item the ACKs must travel back along the reverse path taken by the corresponding data packets,
\item the switch must be able to track the number of ongoing flows; and,
\item the switch must be aware of the window scaling factor of each flow to avoid semantic mismatches between the source and the switch on the receive window svalue.
\end{inparaenum}
Two approaches are possible to achieve the first requirement: either one can implement flow-aware routing in the open source network OS of the bare-metal switches, or, since SDN based switches are more common nowadays, one can rely on the functions already provided by SDN. To fulfill the second requirement, one can implement a SYN/FIN based accounting in hardware registers or in SDN controller to track the number of active flows. 

\textbf{Role of SDN: } In the SDN approach, SDN capability to track flows, flow statistics and the scaling value sent in the SYN segments can be easily invoked to address the three requirements above. In contrast, if SDN is not available and network is not upgrade-able, additional knowledge of the DCN architecture and routing can enable the DCN operator to easily deploy \scheme. For example if single path routing is used, the learning ability of the switches can be invoked to implicitly assume that forward and reverse paths are already the same. If Equal-Cost Multi-Path (ECMP) routing is used a simple modification to the \scheme algorithm to equally divide the flow fair share among the multiple routing paths is easily applied. In addition, to track the number of active TCP flows, we can simply implement an efficient header filters to track SYN/FIN flags for connection establishment or tear-down without per-flow state by using per-port hardware registers.  

\textbf{TCP Window Scaling: } The TCP window scaling option remains an important issue. In practice, this option is supposed to be activated to deal with long-fat pipes by increasing the receiver window from 64KB per flow to up to 1GB per flow. However, even in current low-latency DCN which opted for 10-100 Gbps interfaces, the scaling remains a necessary element and the receive window still needs to be scaled to maintain full link utilization. Even though, one could argue that the chances of having a single flow active on a given port is close to nil (considering the average number of flows per server in a private DCN measurement is 36~\cite{Alizadeh2010}), a robust technique should provide the ability to rescale receive window values. According to the RFC~\cite{RFC1323}, the window scaling option is supposed to be negotiated between the sender and the receiver, and to enable it, both sender and receiver must send the window scaling option in the SYN and its corresponding SYN-ACK. However, in practice the scaling value is not negotiated as different TCP implementations adopt different default values for scaling factor. For example, by default in MacOS the scaling exponent is set to three while Linux calculates it according to the allocated receiver buffer size. Furthermore, these values can be reconfigured by the application to be from 0 (i.e., up to 64KBytes for no-scaling) to 14 (i.e., up to 1 GBytes with scaling). To avoid any cognitive mismatch between the values set by \scheme in the receive window field and those interpreted at the receiver and to operate regardless of the link speed used in modern data centers, the following are the possible workaround tricks to solve this issue:
\begin{itemize}
	\item  If the scaling option is negotiated then we propose to simply unify the value supported in the DCN by rewriting it in the SYN and its corresponding SYN-ACK during the phase of TCP connection establishment via an SDN rule in the switches or directly by \scheme; 
	\item However, if the TCP implementation informs the peers of the scaling value during connection setup, then we propose to have a lightweight shim layer at the end-hosts. This shim-layer tracks the per-flow scaling factor, recomputes the receive window of outgoing ACKs and resets the value to a pre-set network-wide scaling factor which is already pre-known to all the switches. 
	\item The last possibility, which we have adopted in our prototype, is to deploy a module or shim-layer in the hypervisor or server. The module collects the scaling factor used by each side which is typically propagated to the other side with the SYN. Then, the module uses four of the available reserved bits in the TCP header to encode the scaling factor and the switch uses this value to adjust the new receive window whenever it is updated. Hence, the resulting system does not require any changes to TCP and is transparent to VM's guest OS.
\end{itemize}

\textbf{Processing Complexity: } In terms of processing complexity, \scheme is a very simple algorithm with very low complexity and can be integrated easily in switches or routers. For example, it can be implemented in Linux based routers as a module using the NetFilter framework which allows for modifications to the packet headers prior to their forwarding by the IP layer. This requires $O(1)$ computation per packet. \scheme can also cope with Internet checksum recalculation easily and efficiently after header modification, by applying a straightforward one's-complement add and subtract operations on the following three 16-bit words \cite{RFCTCP}: $CSum_{new} = CSum_{old} + Rwnd_{new} - Rwnd_{old}$. In addition, since \scheme is designed to deal with TCP traffic only, tracking the number of flows can be achieved in a scalable manner by monitoring SYN/SYN-ACK and FIN/FIN-ACK bits. This also requires $O(1)$. If necessary, disabling or unifying the window scaling factor in the switch also requires $O(1)$. All in all, all the operations required by the algorithm are $O(1)$ and most importantly involve only the switches/routers under the control of the DC operator. In particular, no modification to the TCP sender or receiver algorithms is needed. 

\textbf{Overhead of \scheme Deployment:} To reduce the overhead which \scheme imposes on core switches, its algorithm can be slightly modified by making use of one of the currently unused three flag bits in the TCP header or the high-order bit of the currently unused fragment offset in the IP header. This bit can be set on a predetermined interval or when the congestion level approaches certain threshold as a flag on the ACKs that were modified by the intermediate routers. Then, when such scheme is applied to \scheme, only the hypervisor, ToR switches and/or lightly-loaded gateway/ingress routers are modified in such a way to store the current up-to-date per-flow receive window value and they become responsible for updating every ACK heading back to the sender. Consequently, by adopting this approach, there is no more extra burden on the intermediate routers or core switches to be involved in updating every ACK passing through them.

\textbf{Effect on Internet-facing TCP connections:} It is worth mentioning that, the WAN connections of data centers to the Internet in most cases are facing intra-Datacenter load balancers and proxies that split the TCP connection. Hence, TCP connections inside the data centers are effectively separated from the TCP traffic from outside the Internet. This avoids the possible issues that \scheme updating the receive window of the Internet-facing TCP connections which runs in a high Bandwidth-Delay Product (BDP) environment like the Internet. 

\section{Convergence and Stability Analysis}
\label{sec:convstab}

Since \scheme adopts proportional increase, proportional decrease approach to adjust its window, it is important to verify its convergence and stability. We can simply model RWQND behavior by considering the three parts that make up the system: per-queue local window updates at the switch, source window adjustments in response to switch feedback (TCP congestion control is assumed to be disabled), and queue behavior. Similar to \cite{Padhye1998, Misra2000}, we adopt a fluid approach to model how \scheme reacts proportionally to the extent of congestion and how it updates the local $Rwnd$ value at a predetermined constant interval in the switch before conveying its value in the incoming ACKs heading to the sources. This leads us to a model that is centered around the switch where all calculations are based on the advances in time. Recall that $T$ is the interval duration of the increments and let $MSS$ be the maximum segment size. We let the target queue occupancy to be $\alpha \times B$ where $B$ is the buffer size and $\alpha$ is the target threshold. At the start of \scheme operation, the local window size $w(t)$ in the switch is initially set to $\alpha \times B$ bytes. We further model the window dynamics using a discrete time model with respect to $T$, then $w(t)$ can be written as:

\begin{equation} 
\label{eq:1} 
  w(t)=\begin{cases}
    w(t-T) + \gamma(t) & \text{if~} t=kMT,\\
    w(t-T) & \text{otherwise},
  \end{cases}
\end{equation}
where, $k$ is a positive integer which tracks the window update epochs and $\gamma(t)$ is the average value of $\kappa(t)$ over the different increment intervals expired during one update interval. Simply put, $\gamma(t)$ is the number of MSS by which the window should be increased or decreased in the update interval preceding $t$. Note that, $\gamma(t)$ is reset at the end of each update epoch (i.e., at time $t=kMT$). Hence, $\gamma(t)$ can be expressed as follows: 

\begin{equation} 
\label{eq:2} 
 \gamma(t)=\begin{cases}
    0 & \text{if~} t=kMT,\\
    \frac{MSS}{M}\left(1-\sum\limits_{j=1}^M{\kappa(t-jT)}\right) = \frac{MSS}{M}\left(1-\sum\limits_{j=1}^M{\frac{q(t-jT)}{\alpha B}}\right) & \text{otherwise}.
  \end{cases}
\end{equation}
Notice that instead of reducing the queue dynamics in the update interval to the final value only, our calculation of $\gamma(t)$ takes into account the past queue fluctuations from the start of the interval by averaging all the sampled $M$ values. Let the link speed (or capacity) be $C$, then the queue dynamics can be described as follows:

\begin{equation} 
\label{eq:3} 
q(t) = \left[q(t-T) + \frac{T}{RTT} w\left(t - \frac{RTT}{2}\right) - CT\right]^+; 
\end{equation}
that is, the queue at time $t$ receives a window full of bytes that was calculated half an RTT earlier (to account for the propagation of the ACK from the switch to the source and the propagation of the data from the source to the switch, arriving at time $t$). For simplicity, we assume there is no congestion and the RTT fluctuates slightly, hence the ACKs do not queue up in the reverse direction.

Then, it is expected that the persistent queue converges to $\alpha \times B$ as $t$ goes to infinity. To support this claim, we use the above model and run numerical experiments in Matlab. In the experiments, we assume traffic sources are connected to one switch with buffer size of 83 packets and the target queue occupancy is set to  $\alpha = 20\% = 16.6$ packets (or pkts). The capacity of the output link is 10 Gbps and the RTT is 100$\mu$s. We run two scenarios where the slow start is enabled in one of them and disabled in the other one.

\begin{figure}[!ht]
\centering
        \includegraphics[height=8cm, width=0.95\textwidth]{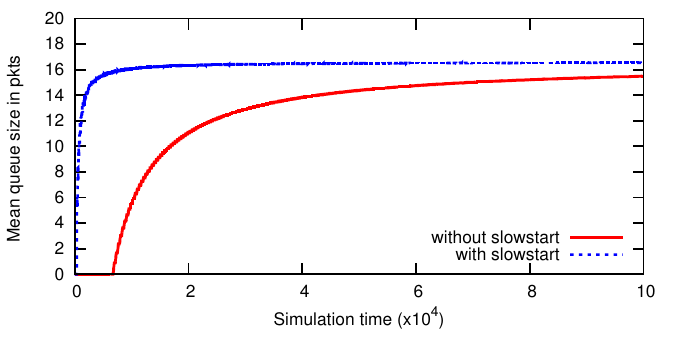}
         \caption{Algorithm stability and convergence speed}
        \label{fig:meanqueuematlab}				
\end{figure}

\figurename~\ref{fig:meanqueuematlab} shows the mean queue size over time for the two scenarios (i.e., with and without slow start) as obtained from running the Matlab simulations. The graph supports our intuition that, at the beginning, the mean queue occupancy stays at the zero level until the pipe is filled. Then, it increases steadily until it converges to the target queue occupancy as time goes to infinity. In addition, slow start seems to improve the speed of convergence dramatically. That is, slow start leads the queue occupancy to the target very fast, then the proportional increase proportional decrease mechanism maintains the window around the target queue occupancy while reacting with agility to congestion. Hence, \scheme enjoys a fast convergence speed and is expected to operate the queue around the target queue level. The system is also expected to be stable as it can reject any external transient disturbance (e.g., new flow or incast event) and return quickly back to the steady state operating point (i.e., the target queue level $\alpha$).

\section{Simulation Analysis}
\label{sec:sim}

In this section, we study the performance of our algorithm via ns2 simulation in network scenarios with a low BDP like data centers. We compare our system to DCTCP and XCP and demonstrate how it outperforms them. We also compared our mechanism to TCP-RED, however we omit their results because its performance was far off the three approaches. 

For \scheme, the values of $\alpha$, $T$ and $M$ are chosen based only on the target level of congestion that can be tolerated regardless of capacity, delay, and number of sources. In the simulation experiments, we set $\alpha$ to 20\% of the buffer size, $T$ to 50 $\mu$s and $M$ to 10 intervals leading to an update epoch every 500 $\mu$s. DCTCP and XCP parameters are set according to their recommended settings with $K$ (the target queue occupancy) of DCTCP set to 17\% of the buffer size. 

\subsection{Single Rooted Tree Topology}

In the following experiments, we use ns2 v2.35 \cite{NS2} which we have extended with the implementation of \scheme as an AQM. In addition, we modified ns2 TCP module because the receiver window interaction between TCP sender and receiver (Flow Control) is not implemented in ns2. We run the experiments for TCP New-Reno with \scheme, DCTCP (which includes a modification of TCP and AQM) and XCP (which is a complete clean-slate approach). For DCTCP, we use a patch for ns2 v2.35 available from the authors \cite{DCTCP} and for proper operation, ECN-bit capability is enabled in the switch and the TCP sender or receiver. For XCP, we use the version available in the ns2 v2.35 \cite{NS2} distribution.

\begin{figure}[!ht]
\captionsetup[subfigure]{justification=centering}
\centering
        \begin{subfigure}[ht]{0.48\columnwidth}
                \includegraphics[height=7cm, width=\textwidth]{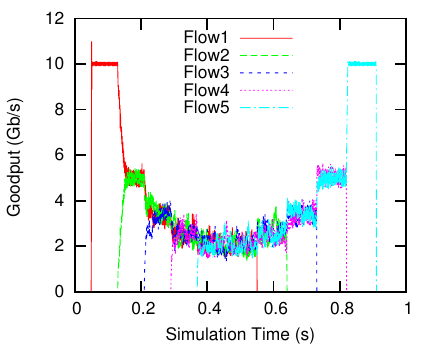}
                \caption{DCTCP}
                \label{fig:dctcp}							
        \end{subfigure}
		\hfill
        \begin{subfigure}[ht]{0.48\columnwidth}
	        \includegraphics[height=7cm, width=\textwidth]{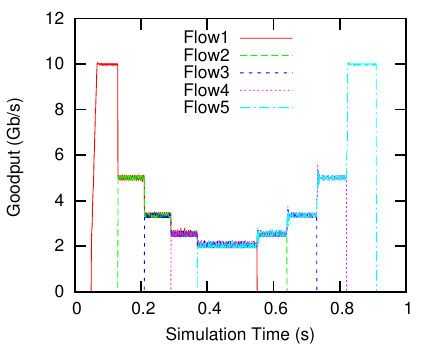}
                \caption{\scheme}
                \label{fig:rwndq}
        \end{subfigure}
	\caption{Goodput of 5 flows that start/stop in a predetermined order showing the convergence speed to the fair-share}\label{fig:fairness}
\end{figure}

\textbf{Simulation Setup: } We use a single rooted-tree (i.e., Dumbbell) topology and run the experiments for a period of 1 sec. The buffer size of the bottleneck link is set to the value of the BDP in all cases (e.g., 83 Packets or 125 KBytes), the IP data packet size is 1500 bytes.  We use in our simulation experiments high speed links of 11 Gbps for the sending stations, a bottleneck link speed of 10 Gbps, a low RTT of 100 $\mu$s and a $RTO_{min}$ of 2 ms as opposed to the default 200 ms. First, we simulate a scenario with 5 elephant flows that start and stop each in a predetermined order to evaluate the convergence and fairness properties of \scheme and DCTCP. Figures~\ref{fig:dctcp} and \ref{fig:rwndq} show the goodput of the 5 elephant flows scenario. The results show that \scheme compared to DCTCP is able to converge faster to the fair-share for each newly active flow and achieves a better short-term (or instantaneous) fairness with a lower variance.

We simulate two other scenarios with 50 and 100 sources, one half of which are elephant flows and the other half are mice flows, to create both the incast and buffer-bloating situations. All sources start at the same time, and while elephants keep sending at full speed during the whole simulation period, mice flows who finish their transfer epoch very quickly, restart sending for another 5 epochs during the simulation. In each of these epochs the different mice flows start in a random order and each flow sends 10 KBytes of data. The interval between the start of two consecutive mice flows is randomly chosen with an average equal to a packet transmission time divided by the number of flows. This allows for the creation of the incast problem where the start times of mice are correlated. We study the CDF of the average and the variance of the Flow Completion Time (FCT) of mice flows, the goodput of elephant flows, the persistent queue size, and the link utilization in the two scenarios. Note that, throughout this thesis, we use both throughput and goodput interchangeably to indicate the rate at which ``useful" data traverses a link.

\begin{figure}[!ht]
\captionsetup[subfigure]{justification=centering}
\centering
        \begin{subfigure}[ht]{0.48\textwidth}
        \includegraphics[height=6cm, width=\textwidth]{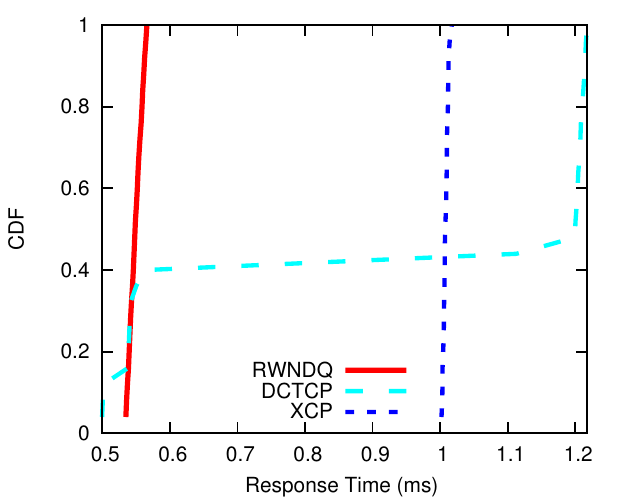}
          \caption{Average FCT of 25 mice flows}
                \label{fig:flow50-cdf}
        \end{subfigure}
        \hfill
        \begin{subfigure}[ht]{0.48\textwidth}
                \includegraphics[height=6cm, width=\textwidth]{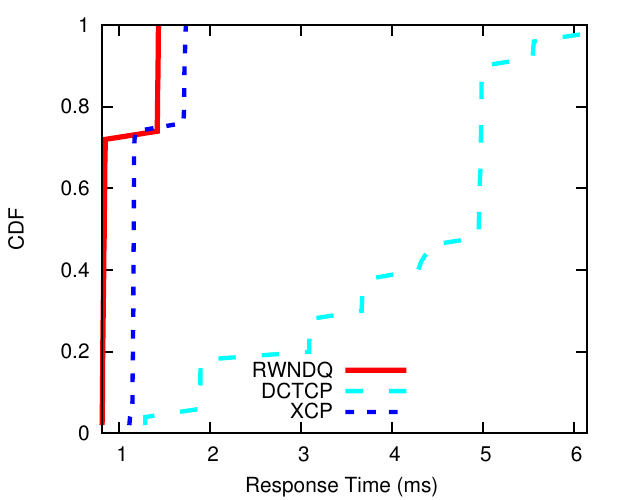}
                \caption{Average FCT of 50 mice flows}
                \label{fig:flow100-cdf}
        \end{subfigure}
		\\
        \begin{subfigure}[ht]{0.48\textwidth}
                \includegraphics[height=6cm, width=\textwidth]{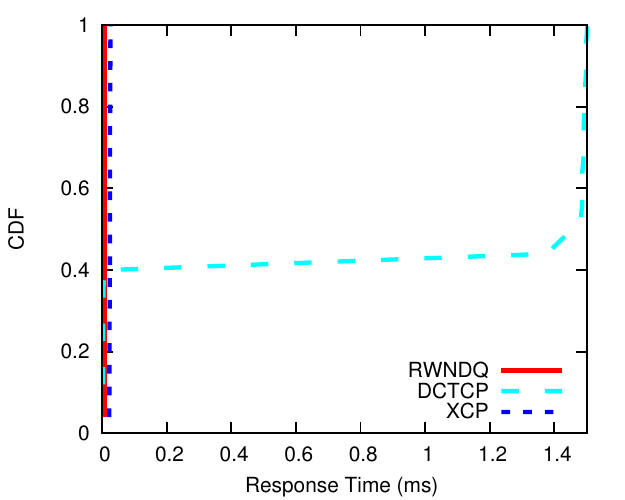}
                \caption{FCT variance of 25 mice flows}
                \label{fig:flow50-varcdf}
        \end{subfigure}
         \hfill
        \begin{subfigure}[ht]{0.48\textwidth}
                \includegraphics[height=6cm, width=\textwidth]{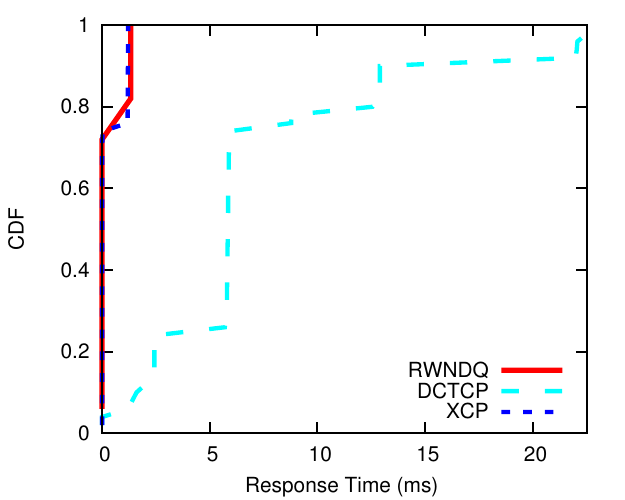}
                \caption{FCT variance of 50 mice flows}
                \label{fig:flow100-varcdf}
        \end{subfigure}
	\caption{The average and variance of FCT for mice flows in 50 and 100 nodes scenarios}
		\label{fig:fct-mean-variance1}
\end{figure}

\begin{figure}[!ht]
\captionsetup[subfigure]{justification=centering}
\centering
        \begin{subfigure}[ht]{0.48\textwidth}
                \includegraphics[height=6cm, width=\textwidth]{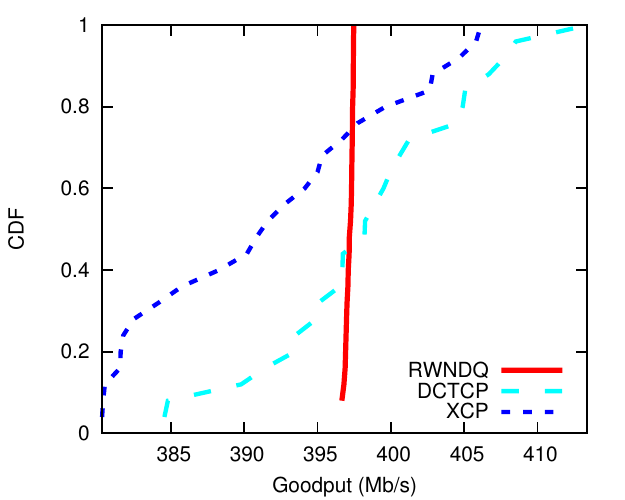}
                \caption{Average goodput of 25 elephants}
                \label{fig:flow50-good}
        \end{subfigure}
        \hfill
        \begin{subfigure}[ht]{0.48\textwidth}
             \includegraphics[height=6cm, width=\textwidth]{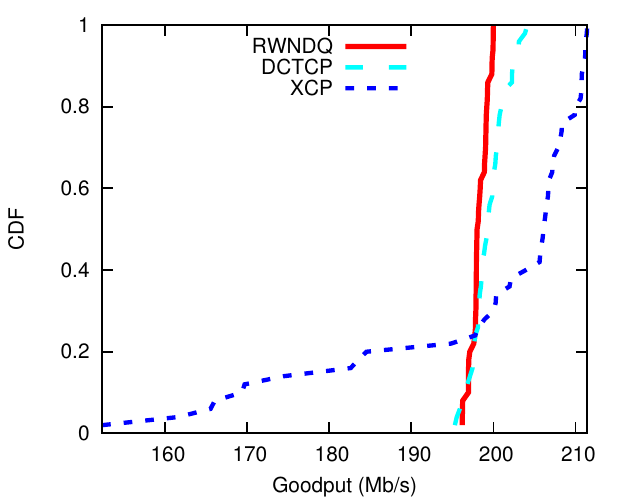}
             \caption{Average goodput of 50 elephants}
             \label{fig:flow100-good}
        \end{subfigure}
	
		\caption{The average goodput for elephants in 50 and 100 nodes scenarios}
		\label{fig:fct-mean-variance2}
\end{figure}

According to Figures~\ref{fig:flow50-cdf}, \ref{fig:flow50-varcdf}, \ref{fig:flow100-cdf} and \ref{fig:flow100-varcdf}, \scheme is able to achieve a faster average FCT compared to DCTCP and XCP and this becomes more conspicuous as the number of flows increases. The variance in the FCT was comparable to that of XCP, \scheme achieves almost zero variance for the 50 flows case and less than 10ms in the 100 flows case; in contrast, for DCTCP the variance reached 33ms in the 100 flows case. Nevertheless, as shown in  Figures~\ref{fig:flow50-good} and \ref{fig:flow100-good} the average goodput of elephants in \scheme is comparable to both DCTCP and XCP with significantly improved fairness among competing elephants. This fairness can be attributed to the faster convergence time of \scheme and the accurate estimation of the congestion level at the switch. By inspecting the number of packet drops, \scheme achieves comparable results to XCP and achieves a lower drop probability than DCTCP. Besides, as the number of flows increases, the number of drops increases further, yet still, \scheme maintains a lower drop probability compared to both DCTCP and XCP. This obviously is the reason for the lower average FCT, as retransmission delay account for the largest portion of high FCTs in our simulations. 

\begin{figure}[!ht]
\captionsetup[subfigure]{justification=centering}
\centering
        \begin{subfigure}[ht]{0.48\textwidth}
               \includegraphics[height=6cm, width=\textwidth]{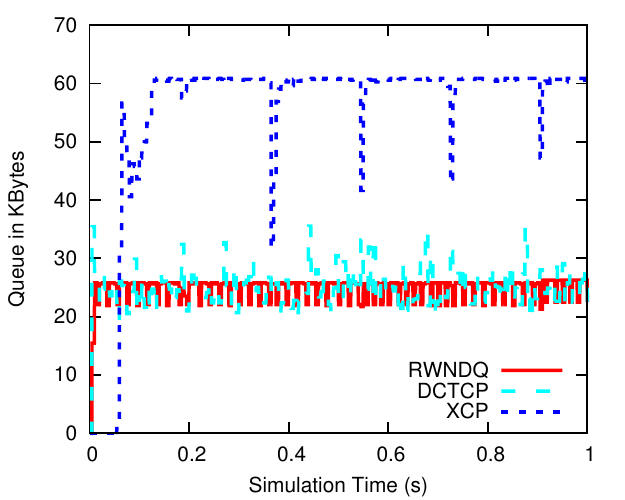}
               \caption{50 flows: persistent queue}
               \label{fig:flow50-persistent}
        \end{subfigure}
        \hfill
        \begin{subfigure}[ht]{0.48\textwidth}
              \includegraphics[height=6cm, width=\textwidth]{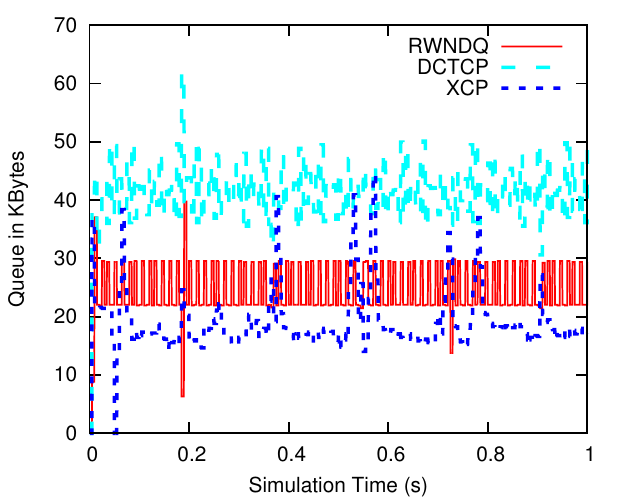}
              \caption{100 flows: persistent queue}
              \label{fig:flow100-persistent}
        \end{subfigure}
  		\\
		 \begin{subfigure}[ht]{0.48\textwidth}
             \includegraphics[height=6cm, width=\textwidth]{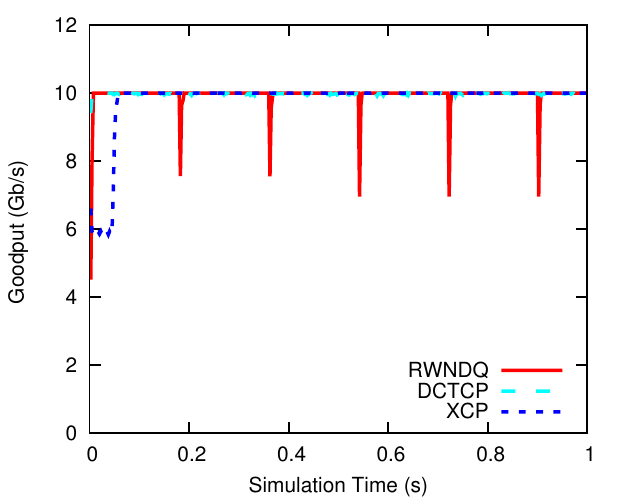}
             \caption{50 flows: link utilization}
             \label{fig:flow50-utilization}
        \end{subfigure}
        \hfill
        \begin{subfigure}[ht]{0.48\textwidth}
             \includegraphics[height=6cm, width=\textwidth]{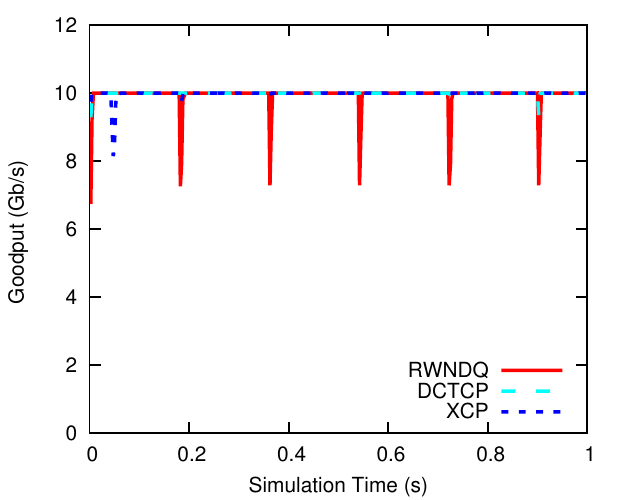}
             \caption{100 flows: link utilization}
             \label{fig:flow100-utilization}
        \end{subfigure}
\caption{The persistent queue and link utilization over experiment time}
		\label{fig:persistent-queue-length}
\end{figure}

Figures~\ref{fig:flow50-persistent} and \ref{fig:flow100-persistent} show the persistent queue length over time. In both cases, the queue stays at nearly the same level for \scheme and always close to the target queue level. This shows the scalability of \scheme as it can nearly stabilize the queue length at the target level even when the traffic volume is high. In contrast for DCTCP, the persistent queue in the 100 flows case is more than double its value in the 50 flows case. However, for XCP, with a larger number of competing flows, the queue occupancy decreases because the feedback values are divided among a larger estimated number of flows which results in fractional increases in the congestion window. The below one fractions eventually get floored at the source to the nearest integer and ultimately the feedback value becomes almost zero.

Figures~\ref{fig:flow50-utilization} and \ref{fig:flow100-utilization} show the bottleneck link utilization. On the one hand, it reveals that DCTCP and XCP can achieve nearly full utilization all the time with little reductions during incast due to their slow convergence (i.e., reaction) time. On the other hand, \scheme has a slightly higher decrease in the utilization at the beginning of incast period. This is due to the fast reaction and fair bandwidth allocation of \scheme when the sudden surge of incast traffic arrive to the queue.

To summarize this simulation study, \scheme seems to be able to smooth oscillations and reach a high link utilization, a small queue size, and a fair rate allocation among competing flows. Besides, the mechanism showed a high degree of robustness in face of varying and sudden traffic surges.

\section{Implementation and Experiments}
\label{sec:exp}

\begin{figure}[ht]
	\centering	 
\includegraphics[width=0.9\textwidth, height=8cm]{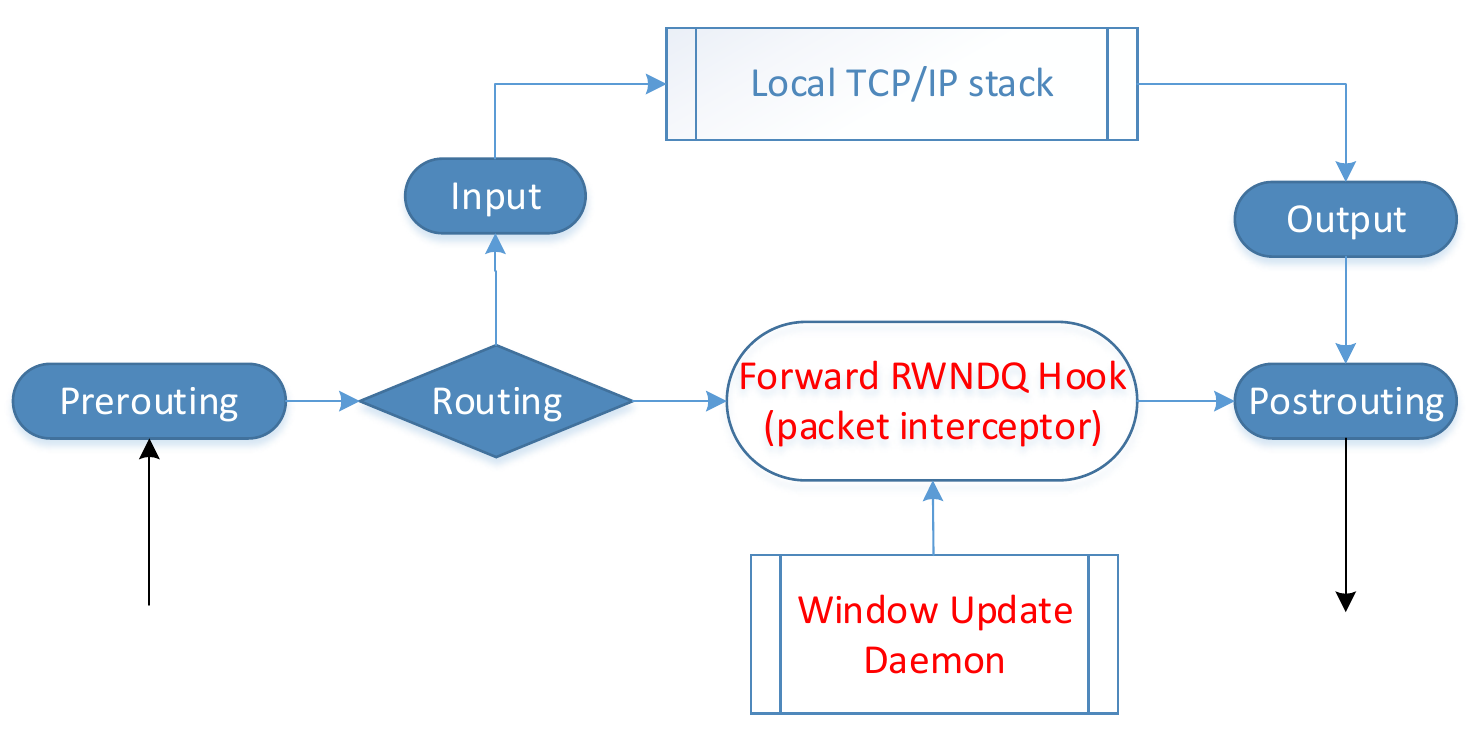}
	\caption{Netfilter-based \scheme packet processing pipeline}
	\label{fig:netfilterpipe}
\end{figure}

We implemented \scheme as a loadable kernel module in the Linux kernel using the NetFilter~\cite{netfilter} framework. In a non-virtualized environment, \scheme module employs hooks that attach to the data forwarding path in the Linux kernel just above the NIC driver and below the TCP/IP stack. This is a clean way of deploying \scheme which does not touch the TCP/IP stack of the software switch. In what follows, we introduce our Linux implementation which is used in the following experiments. 

\figurename~\ref{fig:netfilterpipe} shows the NetFilter hooks structure and interactions in \scheme module. We insert the hook to the forwarding stage of packet processing which intercepts all forwarded TCP packets not destined to the processes running on the server acting as the switch. The forwarding stage is executed right after the routing decision has been made and immediately before the post-routing processing. As explained previously, in the \scheme algorithm, TCP packet headers are examined and the processing is determined based on the SYN-ACK, FIN and/or ACK flag bits. In addition, since in DCNs the transmission and propagation delays are in the microsecond time-scale, Linux kernel timers based on the HZ tick rate (i.e., one jiffy$\approx$1 ms) traditionally used in the protocol stack and  OS, are not fast enough to keep track of the queue occupancy. Therefore, we invoke the high-resolution timers of Linux Kernel to deal with this deficiency\footnote{https://www.kernel.org/doc/Documentation/timers/hrtimers.txt}. The high resolution timer, which is shown in \figurename~\ref{fig:netfilterpipe} as ``Window update daemon'' component, is used to trigger the local per-port window updates. The updates take into account the sampled queue occupancy to more accurately estimate the window. The operations of the \scheme kernel module are described as follows:
\begin{itemize}
\item When a SYN-ACK packet is captured by the NetFilter hook, we increment the connection counter for both the ingress and egress ports and then update their local window register values. SYN-ACK packets is sent only when TCP connection is about to established by the destination host of the connection and by default all TCP connections are full-duplex. Then, assuming the current window value is in a stable steady state and optimal, then a new TCP connection is established or an existing one is deleted, the local window value is redistributed equally and fairly among the active TCP flows.  
 \item When a FIN packet is captured by the NetFilter hook, we only decrement the connection counter for the ingress port and update the local window variable. Even though, FIN packets are only sent when one side of the TCP connection finishes the transport of its application data, the other side of the TCP connection can still send data and the flow end-points operate in a half-closed state until the other side of the connection sends a FIN to its peer.
\item When an outgoing ACK packet is captured by the NetFilter hook, its receive window field of the TCP header is checked against the local window, then the receiver window value is updated only if the local window value is smaller than the receive window in the TCP header. 
\item If the window is updated, the checksum of the packet is recomputed which is done using a built-in kernel function called \textit{"csum\_replace2"} that implements the checksum update in an efficient manner.
\item The high resolution timer is responsible for triggering the local window update function on a regular per-set intervals. Typically, the timer is triggered on intervals less or equal to the dominant RTT (including queueing delays) in the network. For instance, in our setup the RTTs are measured to be within a few hundreds of microseconds without queueing and hence the timer should be set to fire at least once per RTT.
\item For each timer expiry event, we calculate the increment value using the method described in \scheme algorithm~\ref{alg:rwndq}.
\item When the timer expires $M$ consecutive times, the local window value is updated to the accumulated increments value over the past $M$ intervals. Then, the increment variable is reset and a new window update cycle starts at this point.
\end{itemize}

\subsection{Experimental Results and Discussion}

\begin{figure}[!ht]
	\centering	 
\includegraphics[height=7cm, width=0.95\textwidth]{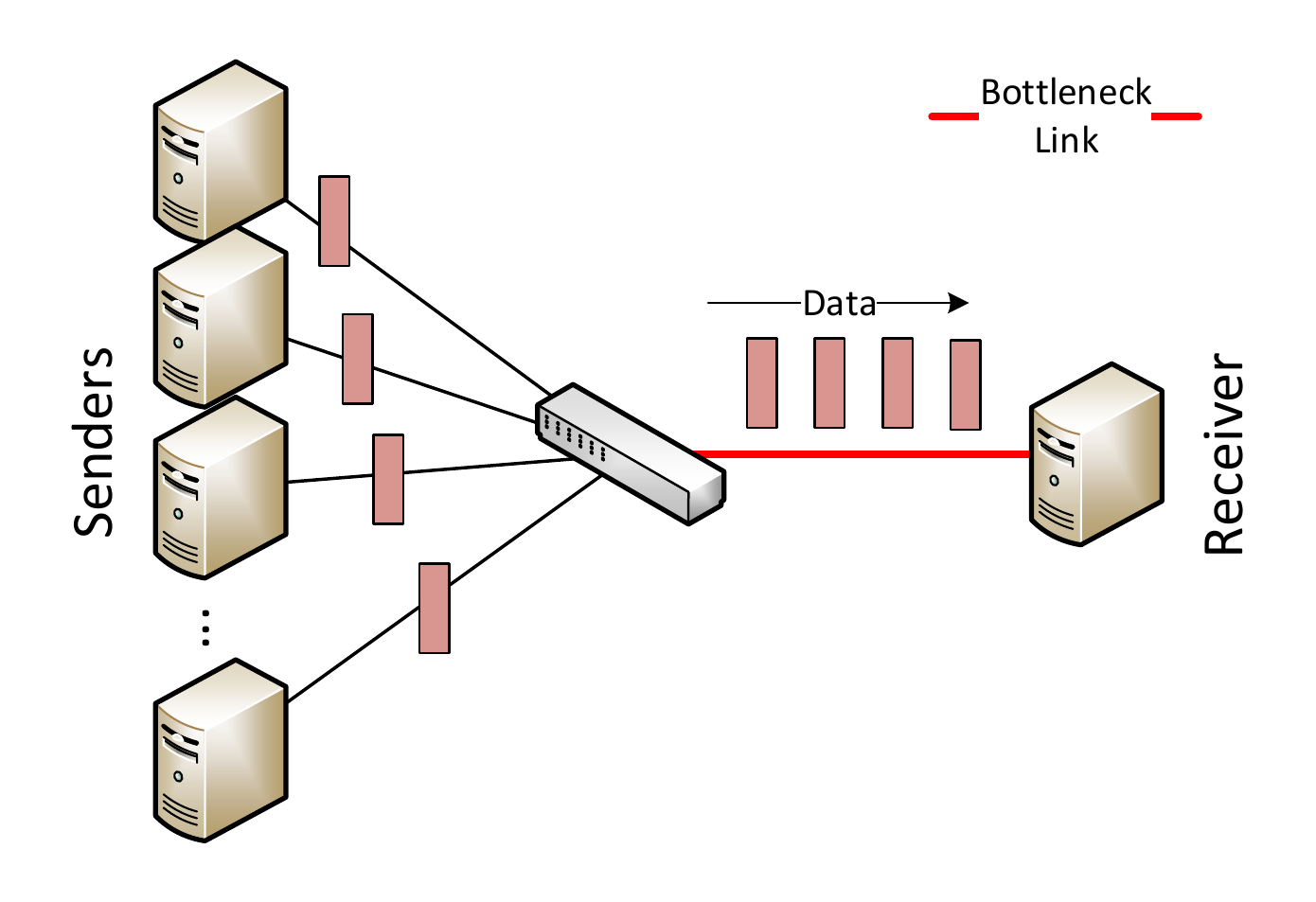}
	\caption{Single-rooted tree testbed to evaluate \scheme as a Linux Kernel Module}
	\label{fig:rwndqtestbed}
\end{figure}

\textbf{Testbed Setup:} To experiment with the Linux \scheme kernel module, we set up a single bottleneck testbed as shown in \figurename~\ref{fig:rwndqtestbed}. The testbed consists of 6 Lenovo and 7 Dell desktops configured with core2duo processors and 4G of RAM. In the setup, one of the Dell desktops is equipped with three 1 Gbps Ethernet cards and acts as the core software switch. Five end-hosts and the switch are running Ubuntu 14.04 with Linux kernel V3.13.0.34, while the other six machines and the receiver (i.e. the master node) are running CentOS 6.6 with Linux kernel V3.10.63. All machines are running an Apache web server which hosts the default \textit{"index.html"} webpage of size 11.5 KB.  

\textbf{Experiment Setup: } In the following experiments, we rely on two well-known measurement applications for our experiments, iperf \cite{iperf} for generating elephants and Apache benchmark \cite{apacheb} for generating synchronized mice. The base RTT in our testbed is around $\approx$200-300$\mu$s without queueing. We allocate a static buffer size of 85.3 KB ($\approx$ 57 full sized packets) to all ports in the network using Linux Traffic Control (Linux TC) system calls. In all experiments, we set up the queuing discipline \textit{Qdisc} of each Ethernet port in the network to \textit{FIFO-DropTail} queue with the aforementioned buffer size. This value matches the per-port buffer size in commodity hardware switches (e.g., Pica8 pronto 3295 bare-metal switch~\cite{Pronto} which has 4MB of buffer space shared by 48 ports that is 85.3 KB of per-port buffering space). We run the experiments with both Cubic and New-Reno (which is abbreviated Reno) TCP, they are the most widely used congestion control mechanisms in Linux kernel as well as other well known operating systems. In the experiments, eleven iperf traffic flows (i.e., elephants) are started at the same time from each of the senders towards the receiver which is connected to one of the core switch ports in \figurename~\ref{fig:rwndqtestbed}. The elephant flows send continuously for 50 seconds and throughput is sampled over 0.5 sec intervals. At the $20^{th}$ second, each sender starts an Apache Benchmark task to request the \textit{"index.html"} page 1000 times (i.e., incast mice traffic) and then the tool reports the statistics on the FCT values. We compare the results of Cubic and Reno with DropTail (DT) vs \scheme as the AQM in the switch.             

\begin{figure}[!ht]
        \centering        
        \begin{subfigure}[ht]{0.48\textwidth}
            \includegraphics[height=6cm, width=\textwidth]{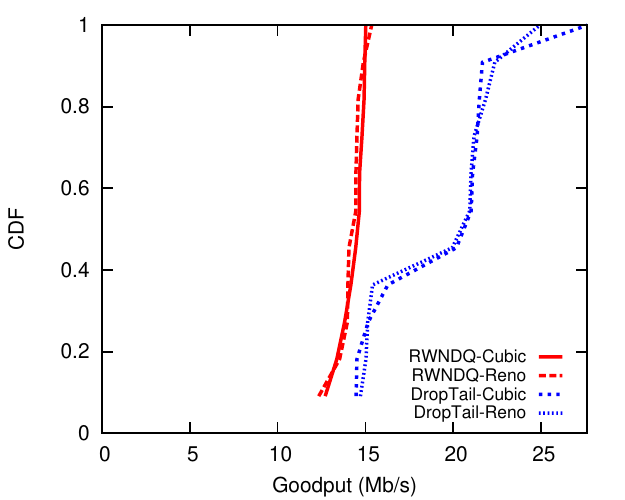}
              \caption{Average elephant goodput}
              \label{fig:module-50-1-goodput-cdf}
        \end{subfigure}
        \hfill
        \begin{subfigure}[ht]{0.48\textwidth}
            \includegraphics[height=6cm, width=\textwidth]{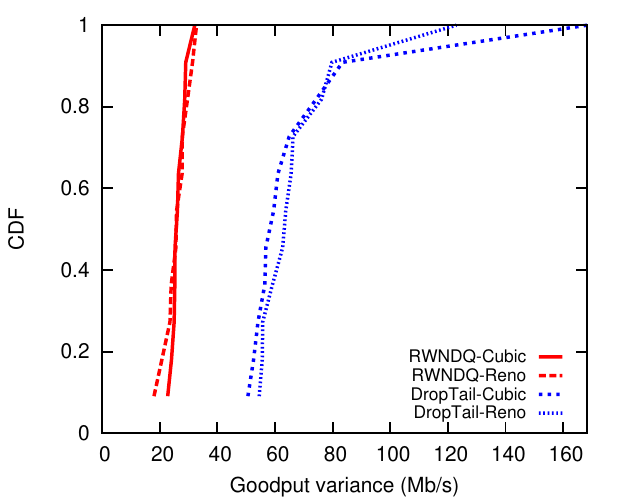}
                \caption{Throughput variance of elephants}
                \label{fig:module-50-1-variance-cdf}
        \end{subfigure}
         \caption{Elephant flows performance: TCP Reno and Cubic with \scheme vs DropTail. The average and variance of the achieved throughput are shown.}
	\label{fig:module-bloat-11senders-good}
\end{figure}

\begin{figure}[!ht]
\captionsetup[subfigure]{justification=centering}
\centering
\begin{subfigure}[ht]{0.48\textwidth}
      \includegraphics[height=6cm, width=\textwidth]{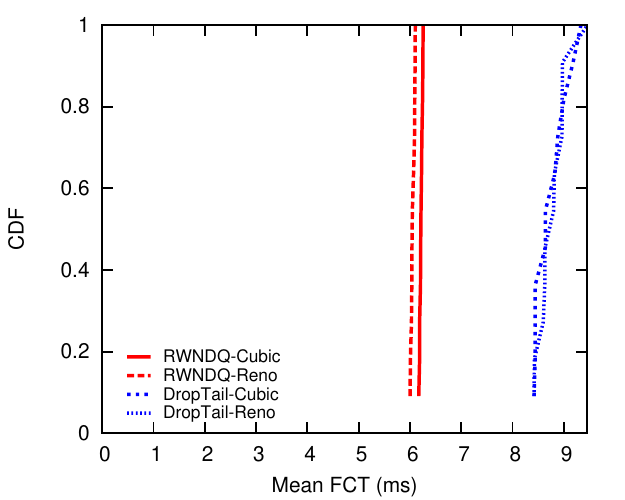}
                \caption{Average mice FCT}
                \label{fig:module-50-1-mean-cdf}
    \end{subfigure}
    \hfill
     \begin{subfigure}[ht]{0.48\textwidth}
      \includegraphics[height=6cm, width=\textwidth]{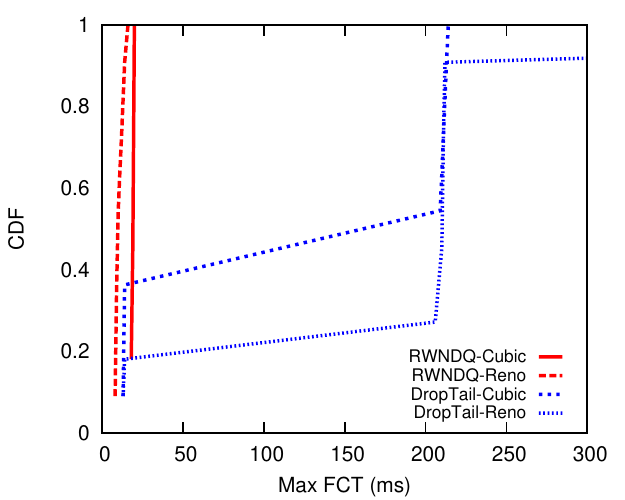}
        \caption{Maximum mice FCT}
         \label{fig:module-50-1-100-cdf}
    \end{subfigure}	
   \caption{Mice flows performance: TCP Reno and Cubic with \scheme vs DropTail. The average FCT and maximum FCT are shown.}
\label{fig:module-bloat-11senders-time}
\end{figure}

\textbf{Experimental Results: } \figurename~\ref{fig:module-50-1-goodput-cdf} shows that \scheme achieves an equal fair-share throughput for the competing elephant flows with both Cubic and Reno . Moreover, \figurename~\ref{fig:module-50-1-variance-cdf} shows that \scheme reduces the variation in the achieved average throughput. Even though, we notice that the average throughput is lower than TCP with DropTail, we relate this to the sudden surges of the large number of synchornized mice traffic (i.e., incast) which enter the network and grab a share equal to the share of the elephants. This is because \scheme gives every flow a fair-share regardless of its type (i.e., mice and elephants are treated equally). \figurename~\ref{fig:module-50-1-mean-cdf} shows that \scheme allows competing mice to finish quickly with little variation in the achieved FCT and most mice flows almost finish their transfer at approximately the same time. More importantly, \figurename~\ref{fig:module-50-1-100-cdf} shows that the tail-end FCT of the mice is improved and that these flows can finish their transfer in less than 200ms. This suggests that mice flows experience very few timeouts whose default minimum value (i.e., $RTO_{Min}$) in Linux is 200ms. By inspecting the packet drop traces (not shown), we conclude that the improvements of \scheme is mainly attributed to the significant reduction ($\approx80-85\%$) in packet drops at the bottleneck link. The reduction of packet drops benefits mostly the mice traffic by avoiding unnecessary timeouts. In summary, \scheme module can give considerable performance gains for mice flows and efficiently redistribute the bandwidth among all competing flows.

\subsection{OpenvSwitch Implementation Details}
\label{sec:exp2}

We also cover the implementation details of \scheme in one of the virtual switches used in current virtualized cloud or data centers. By far, the most well-known and widely used vswitch is the Open vSwitch (OvS)~\cite{OpenvSwitch}. OvS can operate both as a software switch which runs within the end-host hypervisor and as the control stack for the switching silicon in OvS-capable switches (e.g., Pica8 pronto 3259~\cite{Pronto}). It has also been ported to multiple virtualization platforms and switching chipsets. For instance, it is the default switch in XenServer 6.0, the Xen Cloud Platform and also it supports Xen, KVM, Proxmox VE and VirtualBox hypervisors. It has also been integrated into many virtual management systems including OpenStack, OpenQRM, OpenNebula and oVirt. The kernel data-path of OvS is readily distributed with Linux kernel and its user-space installation packages are available for Ubuntu, Debian, and Fedora. Open vSwitch is also supported on FreeBSD and NetBSD operating system. The recent Open vSwitch releases have also been ported to DPDK.

\begin{figure}[!ht]
	\centering	 
	\includegraphics[height=7cm, width=0.9\textwidth]{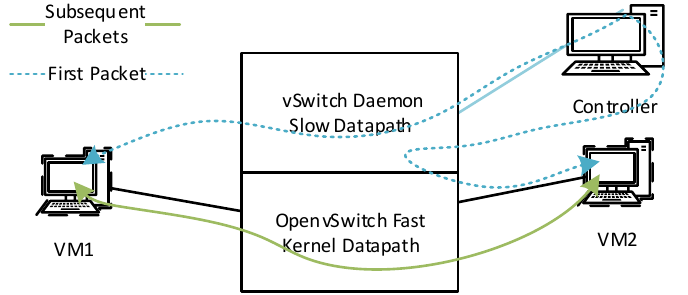}             
	\caption{Open vSwitch slow and fast data-path processing}
   \label{fig:ovsarch}
\end{figure}

We now investigate the implementation of \scheme in OvS. We extended the kernel data-path module of OvS with the same functions described earlier in the \scheme Linux-kernel module. However, in this case, we did not use the NetFilter framework but we have added \scheme functions in the processing pipeline of the packets that pass through the kernel data-path module of OvS. In a virtualized environment, \scheme-enabled OvS can process the traffic for inter-VM, Intra-Host and Inter-Host communications. This is an efficient way of deploying \scheme on the host operating system running the switch by only applying a patch and recompiling the OvS data-path module, making it easily deploy-able in today's production DCs.

\begin{figure}[!ht]
	\centering	 
          \includegraphics[height=8cm, width=0.95\textwidth]{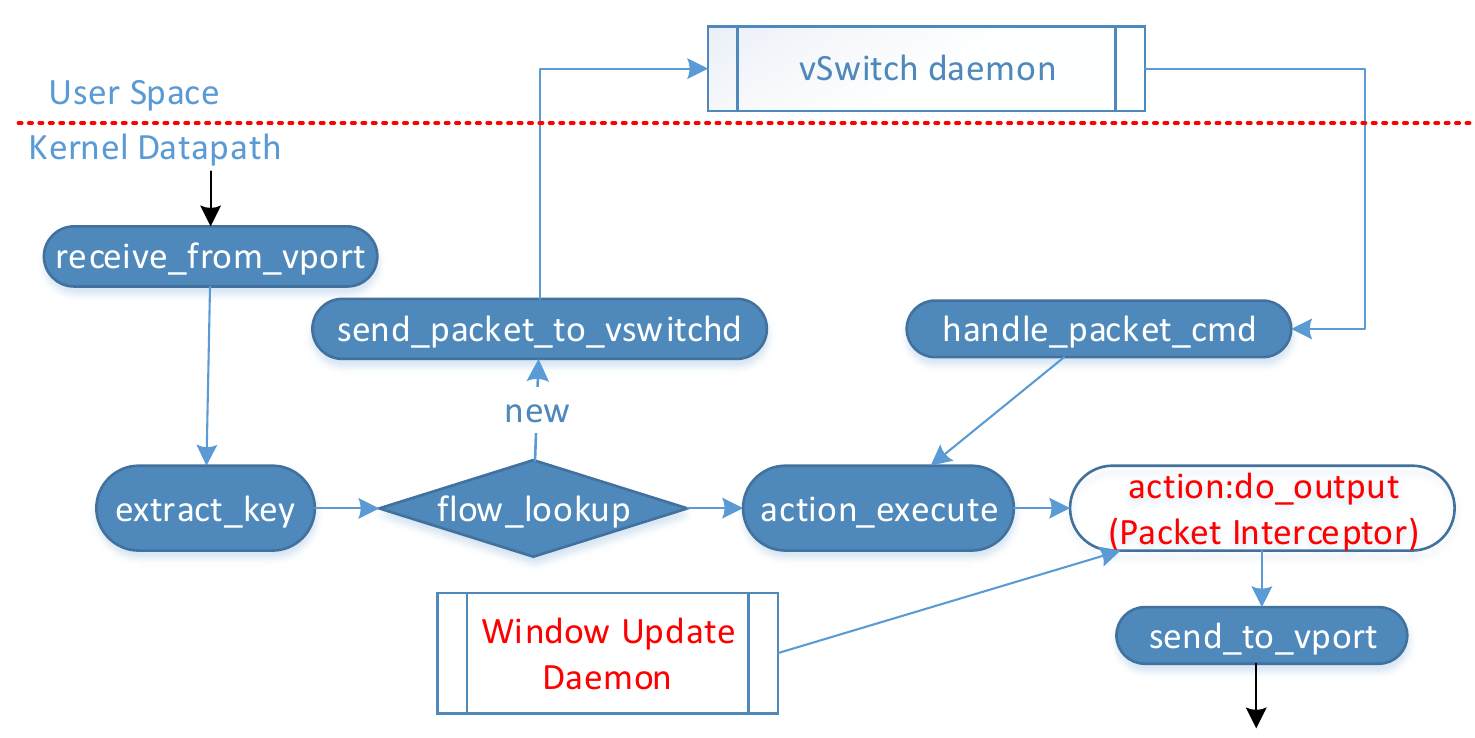}
         \caption{Open vSwitch based \scheme packet processing pipeline}          
         \label{fig:ovspipe}
\end{figure}

As shown in \figurename~\ref{fig:ovsarch}, OvS is mainly composed of two parts, the data-path kernel module and the user-space vSwitch daemon that communicates with the controller using OpenFlow protocol over encrypted SSH connections. OvS is flow-aware by design and all flow decision entries in the forwarding table are inserted by a local or remote controller. Whenever, a packet arrives at any port of the switch, its flow identification key is hashed and examined against the currently active flows in the table. If the entry for that flow could not be found, the packet is immediately forwarded to the controller for making a decision about the fate of this flow (or it will be learnt if it is set as a simple learning switch). After this, the forwarding entries are setup in all the flow tables of the switches involved in the end-to-end data path of the flow. Primarily, any packet is processed by the kernel fast data-path only if its flow entry is active in the forwarding table, in such case, the packet is forwarded immediately without experiencing any further delays. \scheme's packet interceptor and its packet handling logic, described previously in the Linux kernel implementation, are inserted in the processing of $do\_output$ function as shown in \figurename~\ref{fig:ovspipe}. The forwarded TCP packets are intercepted and their window value is updated if necessary. The module also updates the local per-port window values on a regular basis via the window update timeout handler as shown in \figurename~\ref{fig:ovspipe}. 

\subsection{Open vSwitch Experimental Results}

\begin{figure}[!ht]
	\centering	 
\includegraphics[height=7.5cm, width=0.95\textwidth]{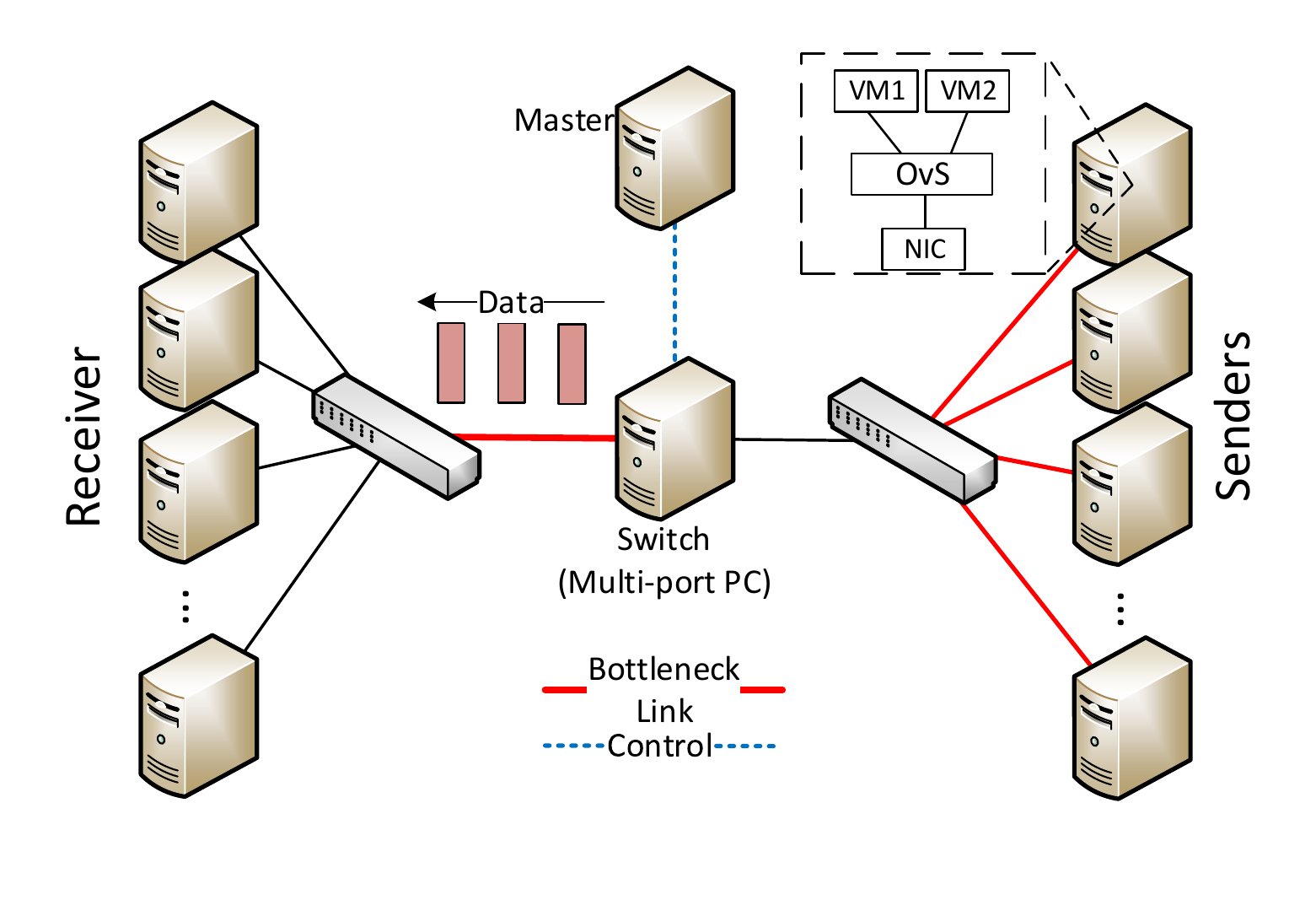}
	\caption{A small multi-rooted tree testbed to evaluate \scheme in Open vSwtich}
	\label{fig:cloudtestbedrwndq}
\end{figure}

For experimenting with our patched OvS, we set up a testbed as shown in \figurename~\ref{fig:cloudtestbedrwndq}, it is similar to the testbed in the previous subsection, however in this case, all the virtual ports and NICs are connected to the patched OvS. The CentOS and Ubuntu end-hosts are again connected to different 1 Gbps D-Link dumb-switches. In this set of experiments, we use different scenarios to produce a mix of incast and buffer-bloating situations, but with multiple bottleneck links present in the network as shown in \figurename~\ref{fig:cloudtestbedrwndq}. The bottlenecks on physical machines are mainly due to the contention between the various internal virtual ports of the OvS which share the same outgoing physical NIC.  

The questions that the next set of experiments answer are: 
\begin{inparaenum}[\itshape i) \upshape]
\item Does the support of \scheme help TCP support many more connections and yet maintain high link utilization? 
\item Does the support of \scheme help TCP overcome incast congestion situations in the network?
\item What is the impact of \scheme on the FCT of mice flows? and,
\item Does \scheme resolve the buffer-bloating situation and manage the buffer efficiently?
\end{inparaenum}

\textbf{Medium Elephant Traffic without Incast: } We produce an a buffer-bloating scenario with senders all converging to the same output port resulting in an excessive pressure on the output buffer of links 1-5 as well as link 6. First, we generate ten iperf clients to start at the same time from each of the five senders which connects to a distinct iperf server listening on a distinct port on the receivers. This results in 50 senders continuously sending for 50 seconds. Iperf is set to collect the throughput samples over 0.5 second intervals. In the following, we show the CDF of the average achieved throughput, the variance of the throughput samples and the total packet drops experienced at the bottleneck links during the experiment. 

\begin{figure}[!ht]
\captionsetup[subfigure]{justification=centering}
\centering
        \begin{subfigure}[ht]{0.48\textwidth}
            \includegraphics[height=6cm, width=\textwidth]{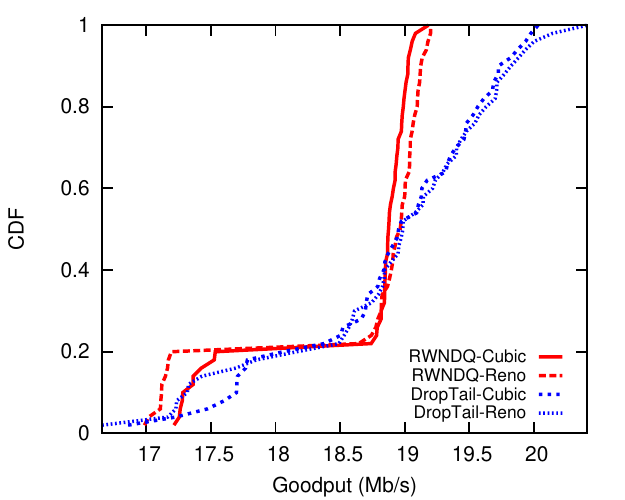}
              \caption{Average elephant goodput}
              \label{fig:incast-50-10-cdf}
        \end{subfigure}
        \hfill
        \begin{subfigure}[ht]{0.48\textwidth}
            \includegraphics[height=6cm, width=\textwidth]{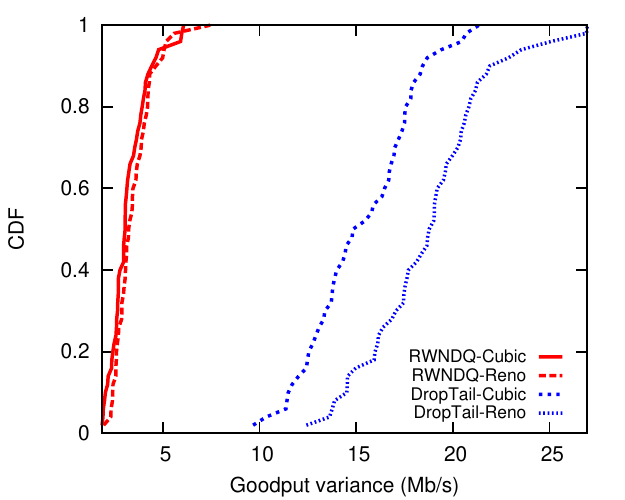}
                \caption{Throughput variance of elephants}
                \label{fig:incast-50-10-varcdf}
        \end{subfigure}
				\caption{TCP elephants performance: TCP Cubic and Reno with \scheme vs DropTail in the 50 elephants experiment}
				\label{fig:incast-10senders}
\end{figure}

\textbf{Experimental Results: }\figurename~\ref{fig:incast-10senders} shows that in a medium load situation, \scheme helps TCP in achieving a balanced distribution of bandwidth among competing senders and reduces the variations of their reclaimed bandwidth during the lifetime of the TCP connection. Table~\ref{tab:10-drops} clearly shows how \scheme is able to reduce the number of packet drops at bottleneck links by $\approx92-99\%$ (nearly two orders of magnitude), hence reducing considerably the number of timeouts for TCP connections and allowing the flows to finish at approximately the same time. This is a result of \scheme's ability to divide the effective window (bandwidth-delay product + target buffer length) equally among competing flows, allowing them to achieve nearly the same throughput and hence almost equal inter-flow FCT. The equal inter-flow completion is an appealing property (esp., for partition-aggregate tasks or co-flows).

\begin{table}[htbp]
\centering
\small
\caption{Number of packet drops observed at each of the 6 bottleneck links labeled [1-6] in \figurename~\ref{fig:cloudtestbedrwndq}}
\begin{subtable}{.47\linewidth}
\centering
\small
\caption{50 elephants scenario}
\begin{tabular}{@{}c|c|c|c|c|@{}}
\cline{2-5}
\multicolumn{1}{c|}{}		& \multicolumn{2}{c|}{Reno} & \multicolumn{2}{c|}{Cubic} \\ \cline{2-5} 
 \multicolumn{1}{c|}{}  & \scheme        & DropTail       & \scheme        & DropTail        \\ \hline
\multicolumn{1}{|c|}{1} & 10           & 4992       & 33            & 4605        \\ \hline
\multicolumn{1}{|c|}{2} & 5            & 4913       & 21           & 4548        \\ \hline
\multicolumn{1}{|c|}{3} & 10           & 4676       & 19            & 4319        \\ \hline
\multicolumn{1}{|c|}{4} & 18           & 4860       & 29           & 4530        \\ \hline
\multicolumn{1}{|c|}{5} & 12           & 4857       & 44            & 4520        \\ \hline
\multicolumn{1}{|c|}{6} & 531          & 331        & 320            & 357         \\ \hline
\end{tabular}
\label{tab:10-drops}
\end{subtable}
\hfill
\begin{subtable}{.47\linewidth}
\centering
\small
\caption{200 elephants scenario}
\begin{tabular}{@{}c|c|c|c|c|@{}}
\cline{2-5} 
 \multicolumn{1}{c|}{}	& \multicolumn{2}{c|}{Reno} & \multicolumn{2}{c|}{Cubic} \\ \cline{2-5} 
 \multicolumn{1}{c|}{}  & \scheme        & DropTail       & \scheme        & DropTail        \\ \hline
\multicolumn{1}{|c|}{1} & 1750           & 30934       & 2184            & 30422        \\ \hline
\multicolumn{1}{|c|}{2} & 1671           & 30851       & 2361           & 30767        \\ \hline
\multicolumn{1}{|c|}{3} & 2544           & 27486       & 2418            & 28276        \\ \hline
\multicolumn{1}{|c|}{4} & 1632           & 30620       & 2152           & 30210        \\ \hline
\multicolumn{1}{|c|}{5} & 1547           & 30860       & 2249            & 30540        \\ \hline
\multicolumn{1}{|c|}{6} & 3394          & 12901        & 3516            & 23432         \\ \hline
\end{tabular}
\label{tab:40-drops}
\end{subtable}
\end{table}
\normalsize

\begin{figure}[!ht]
\captionsetup[subfigure]{justification=centering}
\centering
        \begin{subfigure}[ht]{0.48\textwidth}
            \includegraphics[height=6cm, width=\textwidth]{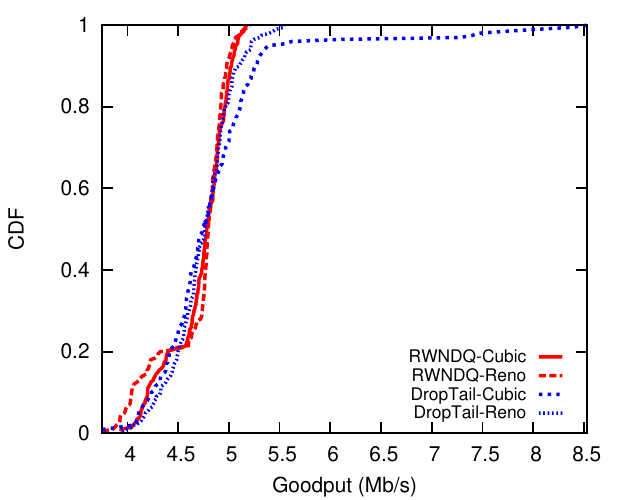}
              \caption{Average elephant goodput}
              \label{fig:incast-50-40-cdf}
        \end{subfigure}
        \hfill
        \begin{subfigure}[ht]{0.48\textwidth}
            \includegraphics[height=6cm, width=\textwidth]{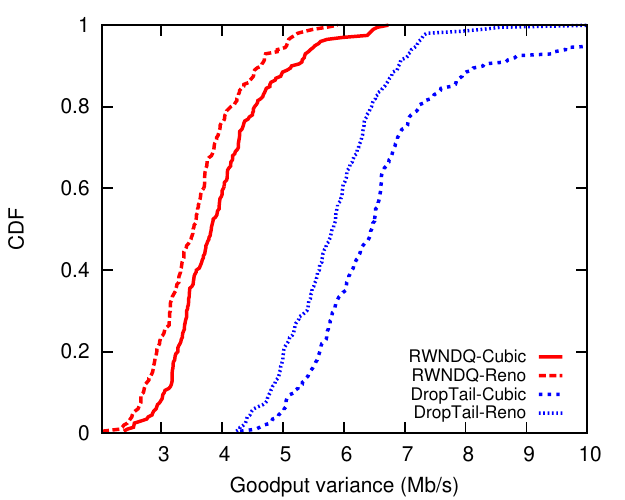}
                \caption{Throughput variance of elephants}
                \label{fig:incast-50-40-varcdf}
        \end{subfigure}
	\caption{TCP elephants performance: TCP Cubic and Reno with \scheme vs DropTail in the 200 elephants experiment}
	\label{fig:incast-40senders}
\end{figure}

\textbf{Heavy Elephant Traffic without Incast: } We repeat the same experiment but this time we increase the number of iperf (elephant) senders per host to 40 (resulting in a total of 200 elephants). Again, \figurename~\ref{fig:incast-40senders} supports our claims, and shows that even in such a high load on the network, \scheme still helps TCP (Reno and Cubic) achieve a balanced distribution of bandwidth and maintains a very low variation of throughput for the TCP connections. We observe that the variation for some TCP flows without \scheme reaches large values of up to $\approx$400Mbps, the reason being, for some time intervals, a few flows grab most of the bandwidth while the others achieve nearly zero throughput. Table~\ref{tab:40-drops} shows that \scheme is still able to keep a very low packet drop rate by one order of magnitude compared to TCP without the assistance of \scheme mechanism at the switch.

\textbf{Medium Elephant Traffic with Incast: } We reproduce the buffer-bloating scenario but with mice traffic competing for the bandwidth with elephant flows to see if \scheme can reconcile the two flow types. Similar to the previous experiment, we first generate ten synchronized iperf elephant connections continuously sending for 50 seconds from each sender resulting in 50 elephants at link 6. We use Apache benchmark to request \textit{"index.html"} webpage (representing mice flows) from each of the web servers ($6\times5=30$ in total) running on the same machines where elephants are sending. Note that, we run Apache benchmark, at the $20^{th}$sec, to request the webpage 1000 times then it reports different statistics over the 1000 requests. 

\begin{figure}[!ht]
\captionsetup[subfigure]{justification=centering}
\centering
   \begin{subfigure}[ht]{0.48\textwidth}
            \includegraphics[height=6cm, width=\textwidth]{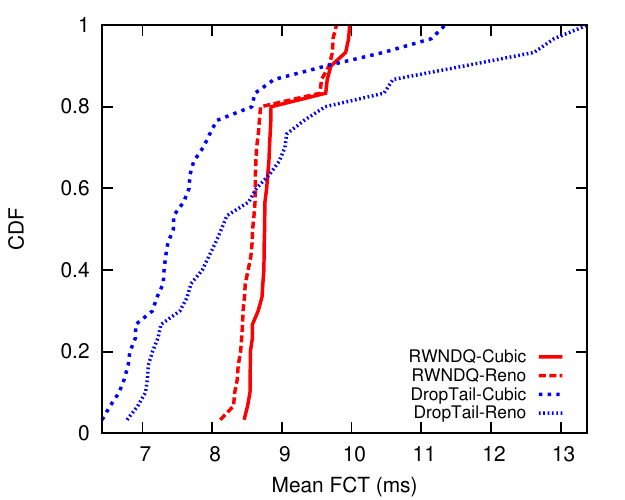}             
		\caption{Average mice FCT}
                \label{fig:bloat-50-10-mean-cdf}
       \end{subfigure}
       \hfill
     \begin{subfigure}[ht]{0.48\textwidth}
            \includegraphics[height=6cm, width=\textwidth]{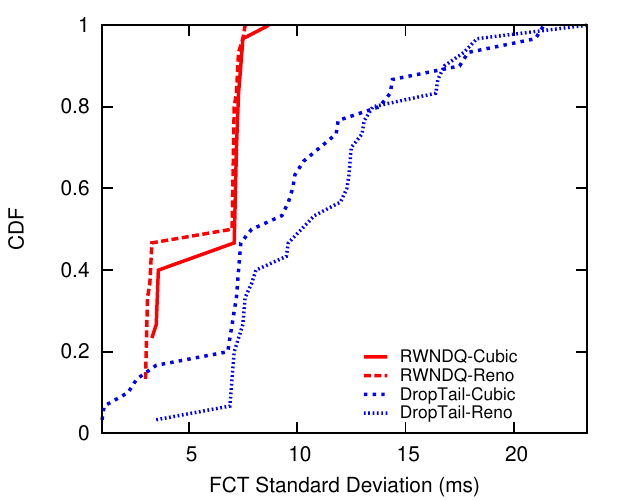}             
	\caption{FCT Standard Deviation for mice}
              \label{fig:bloat-50-10-sd-cdf}
       \end{subfigure}
	\\
    \begin{subfigure}[ht]{0.48\textwidth}
       \includegraphics[height=6cm, width=\textwidth]{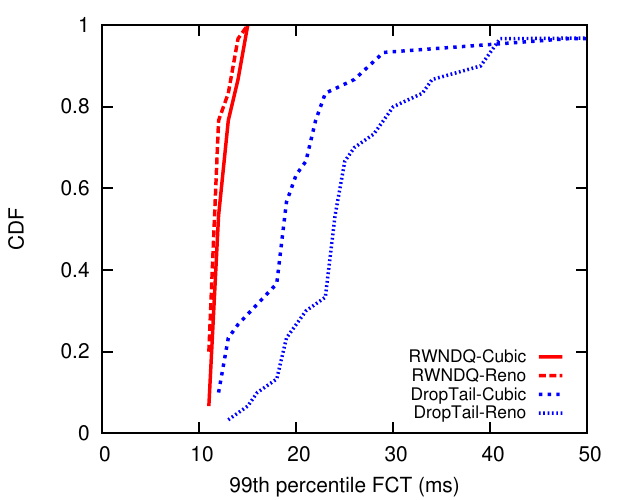}
                 \caption{$99^{th}$-percentile FCT of mice}
                \label{fig:bloat-50-10-99-cdf}
        \end{subfigure}	
        \hfill			
	\begin{subfigure}[ht]{0.48\textwidth}
       \includegraphics[height=6cm, width=\textwidth]{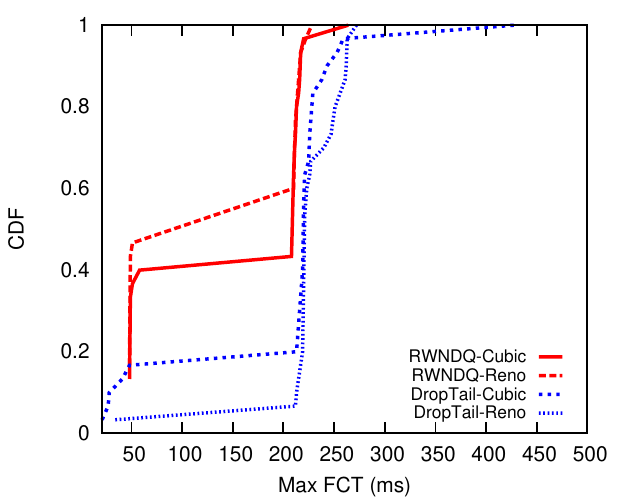}
                 \caption{Maximum mice FCT}
                \label{fig:bloat-50-10-100-cdf}
        \end{subfigure}
	\caption{TCP mice performance: TCP Cubic and Reno with \scheme vs DropTail in 50 elephants against 30 mice}
	\label{fig:bloat-10senders}
\end{figure}

\textbf{Experimental Results: } First, the performance of elephants is not shown because it is close to what has been achieved in the previous scenarios. \figurename~\ref{fig:bloat-10senders} shows that, in medium load, \scheme achieves a good balance in meeting the conflicting requirements of elephants and mice. The competing mice flows benefit under \scheme by achieving a nearly equal FCT on average compared to TCP with DropTail as shown in \figurename~\ref{fig:bloat-50-10-mean-cdf}. \figurename~\ref{fig:bloat-50-10-sd-cdf} shows that the standard deviation is considerably improved with \scheme. In addition, \scheme efficiently regulates the flows and keeps the drop rate near to zero. In \figurename~\ref{fig:bloat-50-10-99-cdf} which shows the $99^{th}$ percentile FCT (i.e., up to 99\% of flows) and in contrast to TCP with DropTail, \scheme never crosses the 200ms threshold which is the default $RTO_{min}$ of Linux. This is can be attributed to the considerably lower number of timeouts thanks to the lower packet drop rates. Moreover, \figurename~\ref{fig:bloat-50-10-100-cdf} shows that the tail-end (maximum) FCT of the ($\approx40-60\%$) flows are below the 200ms with \scheme compared to ($\approx1-18\%$) for TCP with DropTail.

\textbf{Heavy Elephant Traffic with Incast: } Again, we repeat the high load experiment with 200 elephants and introduce 30 competing mice. We justify this large number of elephants because in certain cases the number of long-lived elephants may supersede the mice. For instance, this case can resemble a sudden event (such as a major application failure or large-scale backup) resulting into a large and long-term transfer of data in the network. 

\begin{figure}[!ht]
\captionsetup[subfigure]{justification=centering}
\centering
     \begin{subfigure}[ht]{0.48\textwidth}
        \includegraphics[height=6cm, width=\textwidth]{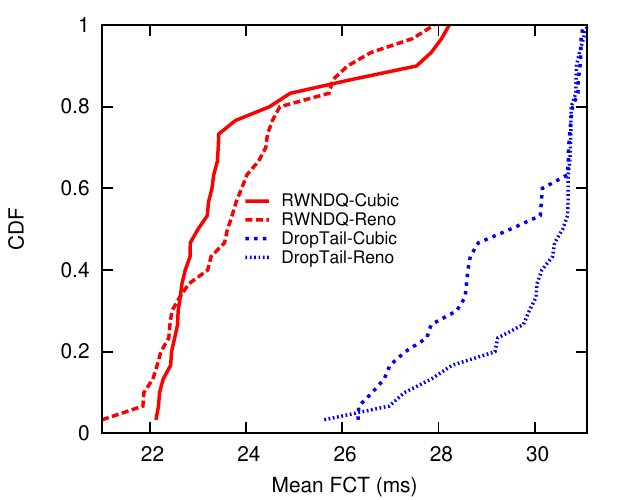}
                \caption{Average mice FCT}
                \label{fig:bloat-50-40-mean-cdf}
        \end{subfigure}
        \hfill
	\begin{subfigure}[ht]{0.48\textwidth}
        \includegraphics[height=6cm, width=\textwidth]{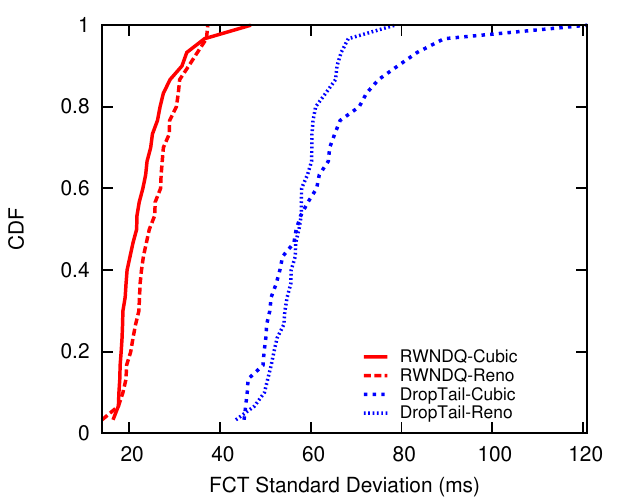}
                \caption{FCT Standard deviation for mice}
                \label{fig:bloat-50-40-sd-cdf}
        \end{subfigure}
	\\
	\begin{subfigure}[ht]{0.48\textwidth}
           \includegraphics[height=6cm, width=\textwidth]{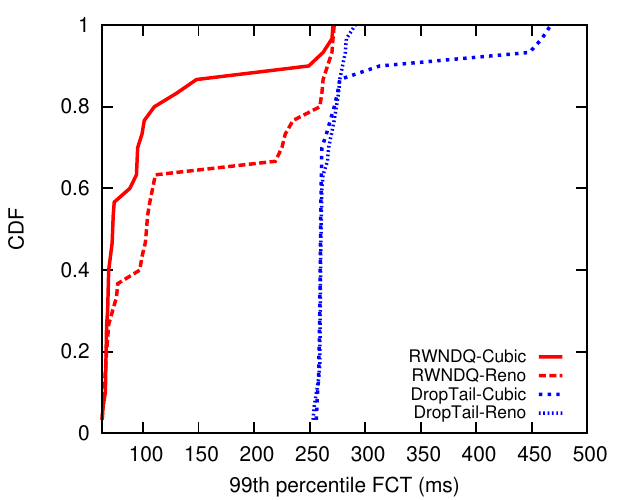}
               \caption{$99^{th}$-percentile FCT of mice}
                \label{fig:bloat-50-40-99-cdf}
        \end{subfigure} 
        \hfill
	\begin{subfigure}[ht]{0.48\textwidth}
           \includegraphics[height=6cm, width=\textwidth]{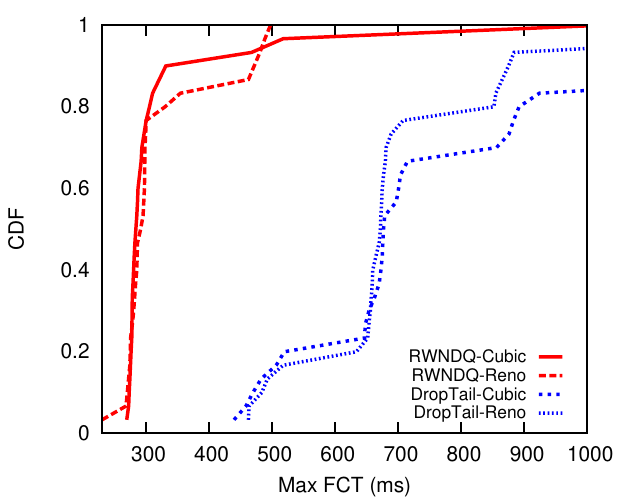}
               \caption{Maximum mice FCT}
                \label{fig:bloat-50-40-100-cdf}
        \end{subfigure} 
	\caption{TCP mice performance: TCP Cubic and Reno with \scheme vs DropTail in 200 elephants against 30 mice}
	\label{fig:bloat-40senders}
\end{figure}

\textbf{Experimental Results: } As shown in \figurename~\ref{fig:bloat-40senders}, \scheme is able to satisfy the requirements of latency-sensitive mice flows, even though they are outnumbered by elephants. In particular, \figurename~\ref{fig:bloat-50-40-mean-cdf} shows that mice flows are not blocked by the bandwidth-hogging elephants. And, \figurename~\ref{fig:bloat-50-40-sd-cdf} shows that FCT variations with \scheme are reduced which is a desirable property for mice flows. The mean FCT under \scheme are small and the CDF curve is also smooth, according to the standard deviation curve, in contrast to what is achieved with DropTail. In addition, \figurename~\ref{fig:bloat-50-40-99-cdf} and~\ref{fig:bloat-50-40-100-cdf} show that, the $99^{th}$ percentile and maximum FCT of tail-end flows with DropTail are too large suggesting that TCP experiences numerous timeouts which can be inferred from the FCT values larger than 250ms. Meanwhile \scheme avoids timeouts by managing the queue efficiently and hence it can greatly reduce the FCT of mice on the $99^{th}$ percentile and tail-end by $\approx60\%$.

\textbf{System Overhead: } To quantify the system overhead introduced by the \scheme packet interception and receive window update function, we measured the CPU usage on the server operating as the switch between the other servers. The server acting as the switch is equipped with Intel Core2 Duo CPU running at 2.13GHz and 4 GBytes of Ram. To measure the overhead, we rerun the high load experiment which involved 200 elephants and the 30 competing mice. We achieved a high link utilization of $\approx$~900-935 Mbps goodput while the extra CPU usage introduced by \scheme is $\approx~1\%$ compared with the case where the \scheme module is not enabled. In summary, \scheme shows a great potential for brining significant performance gains for short-lived flows without incurring severe system overhead.

\section{Related Work}
\label{sec:related}

Due to the impact and severity of the aforementioned congestion symptoms, much recent work has been devoted to addressing such shortcomings of TCP in DCNs. Most of the proposed solutions fall in two categories: window based schemes (e.g., \cite{Ahmed-LCN-2016,Ahmed-ICC-2016-2,Ahmed-LCN-2017,Alizadeh2010,Wu2013, Ahmed-GLOBECOM-2018,Ahmed-hygenicc-techreport2015,Ahmed-GLOBECOM-2015,Ahmed-INFOCOM-2019, Ahmed-ITCE-2019}) or fast loss recovery schemes (e.g., \cite{Vasudevan2009, Cheng2014, Ahmed-INFOCOM-2018, Ahmed-ICDCS-2019-1, Ahmed-ICPP-2020}). In the window-based category, DCTCP \cite{Alizadeh2010} proposes a modification to TCP and RED active queue management that adjusts TCP's congestion window to stabilize the queue length in the switch at a predefined small threshold, guaranteeing thus short delays for incast traffic, without degrading the link utilization. ICTCP \cite{Wu2013} also was proposed as a modification to TCP receiver to handle incast traffic. ICTCP adjusts the TCP receiver window proactively, to avoid congestion at the receiver. The experiments with ICTCP in a real testbed show that ICTCP can almost curb timeout-detected losses and achieves a high throughput for TCP incast traffic, however, since it is focused on incast it only handles congestion at the receiver and does not address buffer buildup in the switches. 
,
Fast loss recovery schemes try to improve the agility of TCP in recovering from congestion events by shortening the reaction time. For instance, \cite{Vasudevan2009,Ahmed-INFOCOM-2018} reduces TCP's minimum retransmission timeout $RTO_{min}$ to reduce the unnecessarily long waiting times after packet losses to enable a fast reaction to congestion losses in the presence of shallow buffers (where losses are mostly detected by timeout). In contrast \cite{Cheng2014} cleverly tries to deploy a fast congestion-detection mechanism by truncating the packet payload of congestion-causing packets, only conveying the header to the receiver. This enables a receiver-driven explicit congestion-notification upon reception of truncated packets. Fast loss recovery schemes potentially solve the problems of congestion in data centers, however, they require not only switch modification, but also end-system modifications. For example, in Linux $RTO_{min}$ is equal to 200ms and is hard-coded in the TCP source code. Other approaches consider the co-flow abstraction to collectively optimize the performance of flows who share the same goal or task~\cite{Ahmed-ICC-2019,Ahmed-SRDS-2017,Ahmed-SRDS-2017,Ahmed-ICDCS-2019-2}.

Alternatively, we have explored an end-to-end flow-aware approach~\cite{Ahmed-CLOUDNET-2015,Ahmed-IPCCC-2015} leveraging explicit feedback from in-network devices similar to traditional flow-based systems like ATM-ABR~\cite{ATM-ABR} or XCP~\cite{Katabi2002}. We have addressed a practically difficult challenge that arises which is how to deploy such flow-awareness in the flow-averse IP environment without modifying the TCP sender and receiver. This disqualifies XCP, as it is a clean-slate redesign that requires not only changes to the routers but also to the sender and receiver. To achieve our goal, the switch/router is set to track flows, calculate a fair share for each flow that traverses it, and convey back this fair share to the source. In our approach, we enable low profile flow awareness, and modify the switch software to rewrite TCP receiver window to communicate with the sender. Hence, our proposed mechanism fits in well for data centers without any change to TCP at the end-hosts. These works were followed-up with SDN-based designs which leverages the same idea towards solving the problem~\cite{Ahmed-ICC-2016-1,Ahmed-sicc-techreport2016,Ahmed-sdngcc-techreport2016,Ahmed-LCN-2017,Ahmed-ANNALS-2017}

\section{Conclusion}
\label{sec:conclude}
In this work, we explore a non-intrusive way of reconciling between the long-lived (elephants) and short-lived (mices) flows. To achieve this, the persistent switch queue sizes should operate at low levels to make room for the bursts of incast traffic which helps avoid packet-losses. We proposed, \scheme, a switch-assisted flow-aware rate matching algorithm that only relies on the existing flow-control mechanism of TCP to feedback queue occupancy levels to TCP senders. To show the practicality of \scheme, we provide prototypes implemented as Linux-Kernel module for bare-metal switches, OpenvSwitch patch for virtual DC networks in data centers. A number of detailed simulations and real test-bed experiments showed that \scheme can achieve its goals efficiently while outperforming the most prominent alternative approaches. Last but not least, knowing that in most public data centers the TCP sender and/or receiver are outside the control of the DCN operator, \scheme makes a point of principle to not modify the source TCP congestion control algorithms to enable true deployment potential in real public DC networks.

\bibliographystyle{IEEEtran}
\bibliography{main}

\end{document}